\documentclass[aps,prd,twocolumn,amsmath,superscriptaddress,amssymb,showpacs,floatfix,nofootinbib,longbibliography]{revtex4-1}

\usepackage{graphicx}
\usepackage{dcolumn}
\usepackage{xcolor}
\usepackage{bm}
\usepackage{natbib}
\usepackage{calligra}
\usepackage[T1]{fontenc}
\usepackage{egothic}
\usepackage[T1]{fontenc}
\newfont{\rsfsten}{rsfs10 scaled 1200}
\newfont{\rsfsseven}{rsfs10 scaled 1200}
\newfont{\rsfsfive}{rsfs10 scaled 1200}
\usepackage{epsfig}
\usepackage{units}
\usepackage[utf8]{inputenc}
\usepackage{hyperref}

\newcommand{\aap}{{Astron.~Astrophys.}}
\newcommand{\apjl}{{Astrophys.~J.~Lett.}}
\newcommand{\apjs}{{Astrophys.~J.~Supp.}}
\newcommand{\aj}{{Astron.~J.}}
\newcommand{\mnras}{{Mon.~Not.~R.~Astron.~Soc.}}

\newcommand{\ssr}{{Space~Science~Rev.}}

\newcommand{\be}{\begin{equation}}
\newcommand{\ee}{\end{equation}}
\newcommand{\bea}{\begin{eqnarray}}
\newcommand{\eea}{\end{eqnarray}}

\newcommand{\GeV}{{\rm\; GeV}}

\def\lsim{\mathrel{\raise.3ex\hbox{$<$\kern-.75em\lower1ex\hbox{$\sim$}}}}
\def\gsim{\mathrel{\raise.3ex\hbox{$>$\kern-.75em\lower1ex\hbox{$\sim$}}}}

\begin{document}

\title{Utilizing cosmic-ray positron and electron observations to probe the averaged properties of Milky Way pulsars}

\author{Ilias Cholis}
\email{cholis@oakland.edu, ORCID: orcid.org/0000-0002-3805-6478}
\affiliation{Department of Physics, Oakland University, Rochester, Michigan, 48309, USA}
\author{Iason Krommydas}
\email{iason.krom@gmail.com}
\affiliation{Physics Division, National Technical University of Athens, Zografou, Athens, 15780, Greece}
\date{\today}

\begin{abstract}
Pulsars have long been studied in the electromagnetic spectrum. Their environments are rich 
in high-energy cosmic-ray electrons and positrons likely enriching the interstellar medium with such 
particles. In this work we use recent cosmic-ray observations from the \textit{AMS-02}, \textit{CALET} 
and \textit{DAMPE} collaborations to study the averaged properties of the local Milky Way pulsar 
population.

We perform simulations of the local Milky Way pulsar population, for interstellar medium assumptions 
in agreement with a range of cosmic-ray nuclei measurements. Each such simulation contains $\sim 10^{4}$ 
pulsars of unique age, location, initial spin-down power and cosmic-ray electron/positron spectra. We produce more
than $7\times 10^{3}$ such Milky Way pulsar simulations. We account for and study i) the pulsars'  birth rates 
and the stochastic nature of their birth, ii)
their initial spin-down power distribution, iii) their time evolution in terms of their braking index and 
characteristic spin-down timescale, iv) the fraction of spin-down power going to cosmic-ray electrons and positrons and 
v) their propagation through the interstellar medium and the Heliosphere. We find that pulsars of ages $\sim 10^{5}-10^{7}$ yr, 
have a braking index that on average has to be 3 or larger. Given that electromagnetic spectrum observations of young pulsars 
find braking indices lower than 3, our work provides strong hints that pulsars' braking index increases on average as they age, allowing them  
to retain some of their rotational energy. Moreover, we find that pulsars have relatively uniform properties as sources of 
cosmic-ray electrons and positrons in terms of the spectra they produce and likely release O($10\%$) of their 
rotational energy to cosmic-rays in the ISM. Finally, we find at $\simeq$12 GeV positrons a spectral feature that 
suggests a new subpopulation of positron sources contributing at these energies. 
\end{abstract}

\maketitle

\section{Introduction}
\label{sec:introduction}
Pulsars represent a class of energetic sources whose properties have been probed over 
more than 50 years via observations in the electromagnetic spectrum. Emission from 
pulsars and their environments has been detected in the radio, \cite{1968Natur.217..709H, 
1998MNRAS.301..235G, 1999ApJS..123..627S, Weisberg_1999, Everett:2000yj, 
McLaughlin:2003zz, Weisberg:2003ud, 2005AJ....129.1993M}, infrared and visible 
\cite{2009A&A...504..525S, 2012A&A...544A.100M, 2012A&A...540A..28D, 2013MNRAS.433.3325S, 
2019A&A...629A.140S}, ultraviolet \cite{2019ApJ...871..246M, 2021NatAs...5..552A}, 
X-rays,\cite{1971ApJ...167L..67G, Becker:2002wf, Gentile:2013yka, 2013MNRAS.433.3325S, 
Guillot:2019vqp, Zhao:2021ilq}, gamma-rays \cite{1994ApJS...90..789U, 2010ApJS..187..460A, 
Fermi-LAT:2013svs, Buhler:2013zrp, Cholis:2014noa, Fermi-LAT:2019yla} and most recently, 
a clear detection of powerful Milky Way pulsars at $O(10)$ TeV gamma-rays has been established 
\cite{2009ApJ...700L.127A, MAGIC:2015ggt, HESS:2017lee, Abeysekara:2017hyn, HAWC:2019tcx}. 
Most of the observed photons from pulsars and their surrounding pulsar wind nebulae (PWNe) 
-where those are present- originate from cosmic-ray electrons and positrons and are emitted via 
curvature radiation \cite{1990A&A...234..237G, 2013IAUS..291..552W}, synchrotron radiation 
\cite{DelZanna:2006wj, Buhler:2013zrp, Porth:2013boa} and at the highest energies inverse 
Compton emission \cite{1996MNRAS.278..525A, 2010A&A...523A...2M, Buhler:2013zrp, 
Kargaltsev:2015pma}. The fact that we have observed $O(10)$ TeV gamma-rays from certain 
pulsars that are still surrounded by their respective PWN clearly sets a lower limit on the electron 
and positron cosmic-ray energies in these environments.   We expect that such pulsars will act 
as sources of cosmic-ray electrons and positrons that are released into the interstellar medium 
(ISM). In fact, we expect for electrons and positrons to be further accelerated as they propagate 
through the termination shock of the respective PWNe before entering the ISM \cite{1996SSRv...75..235A, 
Malyshev:2009tw}. If Milky Way pulsars are prominent sources of high-energy cosmic-ray electrons 
and positrons then we could expect to see their contribution to the relevant cosmic-ray measurements 
and most notably in the cosmic-ray positron flux spectrum.

Cosmic-ray positrons are produced in inelastic collisions of high-energy cosmic-ray nuclei with the 
ISM gas and are typically referred to as secondary positrons. In the same type of interactions matter 
cosmic-ray secondary electrons and secondary nuclei as Boron are produced.  Those have been modeled in 
\cite{Moskalenko:2001ya, Kachelriess:2015wpa, GALPROPSite, Strong:2015zva, Evoli:2008dv, 
DRAGONweb, Evoli:2011id, Pato:2010ih, Cholis:2021inprep} and are in agreement with the current 
observations \cite{Aguilar:2015ctt,  Aguilar:2017hno, Aguilar:2018njt}. A prominent exception is the 
spectrum of the positron fraction $e^{+}/(e^{+} + e^{-})$, measured by \cite{PAMELA:2013vxg, 
AMS:2014bun, AMS:2019iwo, AMS:2021nhj} to rise above 5 GeV,  in disagreement with the expectation from same 
type of models. This suggests an additional source of high-energy cosmic-ray positrons.  Such 
positrons can come from near-by Milky Way pulsars \cite{1987ICRC....2...92H, 1995PhRvD..52.3265A, 
1995A&A...294L..41A, Hooper:2008kg, Yuksel:2008rf, Profumo:2008ms, Malyshev:2009tw, 
Kawanaka:2009dk, Grasso:2009ma, Linden:2013mqa, Cholis:2013psa, Yuan:2013eja, Yin:2013vaa, 
Cholis:2018izy, Evoli:2020szd, Manconi:2021xom, Orusa:2021tts}.  One alternative to pulsars includes local and 
recent supernova remnants (SNRs) \cite{Blasi:2009hv, Mertsch:2009ph, Ahlers:2009ae, Blasi:2009bd, 
Kawanaka:2010uj, Fujita:2009wk, Cholis:2013lwa, Mertsch:2014poa, DiMauro:2014iia, Kohri:2015mga, 
Mertsch:2018bqd}. However, given that SNRs are the major source of all cosmic rays, in order to explain 
the rising positron fraction, the metallicities of environments of recent and close-by SNRs have to be 
different from those of the Milky Way on average \cite{Cholis:2013lwa, Mertsch:2014poa, Cholis:2017qlb, 
Tomassetti:2017izg}. Another possibility is that of particle dark matter \cite{Bergstrom:2008gr, Cirelli:2008jk, 
Cholis:2008hb, Cirelli:2008pk, Nelson:2008hj, ArkaniHamed:2008qn, Cholis:2008qq, Cholis:2008wq, 
Harnik:2008uu, Fox:2008kb, Pospelov:2008jd, MarchRussell:2008tu, Chang:2011xn, Cholis:2013psa,  
Dienes:2013xff, Finkbeiner:2007kk, Kopp:2013eka, Dev:2013hka, Klasen:2015uma, Yuan:2018rys, 
Sun:2020dla}. While such particle dark matter models have been constrained by cosmic-microwave 
background (CMB) data \cite{Slatyer:2009yq, Evoli:2012zz, Madhavacheril:2013cna,  Ade:2015xua, 
Slatyer:2015jla, Poulin:2016nat} and $\gamma$-rays \cite{Tavakoli:2013zva, Geringer-Sameth:2014qqa, 
Ackermann:2015zua}, the full parameter space has not been entirely excluded. 

In this paper we are going to use the cosmic-ray observations from the Alpha Magnetic Spectrometer 
(\textit{AMS-02}), the Calorimetric Electron Telescope (\textit{CALET}) onboard the International Space 
Station and the DArk Matter Particle Explorer (\textit{DAMPE}) satellite. We are going to set constraints 
on the contribution of local Milky Way pulsars to the electron and positron ($e^{\pm}$) fluxes and in turn 
probe their averaged properties. For cosmic-ray energies $E$ above $10$ GeV the propagation of 
electrons and positrons is mostly affected by their energy losses due to synchrotron radiation and inverse 
Compton scattering \cite{RevModPhys.42.237}. The relevant energy loss timescale for electrons and 
positrons of initial energy $E_{\textrm{init}}$ to lose half its energy is,
\begin{equation}
\tau_{\textrm{loss}}(E_{\textrm{init}}) \simeq 20 \times \left( \frac{E_{\textrm{init}}}{10 \textrm{GeV}} \right)^{-1} 
\; \textrm{Myr}.
\label{eq:E_loss}
\end{equation}
Pulsars lose their rotational kinetic energy within $O(10)$ kyr. Roughly that is also equal to the time that 
magnetic fields in the surrounding PWNe and the further out SNR become weak enough to allow the 
relevant cosmic-ray $e^{\pm}$ to effectively escape. The $O(10)$ kyr timescale is typically one to four orders 
of magnitude smaller by comparison to the timescale cosmic-rays need to reach us via diffusion 
from $O(100)$ pc - $O(1)$ kpc distances where most pulsars are at. Thus pulsars can be treated as approximately 
injecting an appreciable fraction of their rotational energy to cosmic-ray electrons and positrons at the 
beginning of their existence \cite{Profumo:2008ms, Malyshev:2009tw} \footnote{A recent work on pulsars surrounded 
by TeV halos, has suggested that some pulsars may be able to contain their cosmic rays well after their birth 
\cite{Mukhopadhyay:2021dyh}. As long as the injection phase of GeV-TeV cosmic rays from these environments 
into the ISM occurs on a timescale significantly smaller than their propagation time, our basic analysis is not 
changed as this would be on average an overall time-shift for the pulsar population.}. 

A result of Eq.~\ref{eq:E_loss}, is that as the observed energy of cosmic rays increases, the number of potential 
sources drops given that only most recent pulsars have an age that is similar to $\tau_{\textrm{loss}}$.  As 
pulsars are born in the Milky Way with a rate of $\simeq$1 per century \cite{1999MNRAS.302..693D, 
Vranesevic:2003tp, FaucherGiguere:2005ny, Lorimer:2006qs, Keane:2008jj}, only a small number of them 
can contribute, and only from an increasingly smaller distance. The relation connecting the maximum energy 
that electrons and positrons can have originating from a distance $R$ was approximated in \cite{Cholis:2017ccs} to be 
$E_{\rm max} \sim 100\, {\rm GeV} (R/2\,{\rm kpc})^{-2}$. For instance at 500 GeV we are probing only pulsars 
from within $\sim$ 400 pc. As there is only a small number of such pulsars, the discreteness of those sources 
will result in subsequent features \cite{Profumo:2008ms, Malyshev:2009tw, Cholis:2017ccs, Fornieri:2019ddi} 
(see also \cite{Mertsch:2018bqd} for a similar study on the impact of recent SNRs). With the recent refined 
observations by \textit{AMS-02} and the observations by \textit{CALET} and \textit{DAMPE} that extend up to 
5 TeV we  will probe the properties of these pulsars. The lower energies of 5-500 GeV, are also used and 
provide us with valuable information on the averaged properties of pulsars that are now "middle aged" and of 
up to O(10) Myr and are located within 4 kpc. Finally, we will show that we can also asses information 
on the properties of the ISM within that same volume.

In section~\ref{sec:Modeling} we discuss the simulations that we perform to account for astrophysical 
uncertainties on a) the stochastic birth distribution in space and time of the pulsar source-population, b) the 
initial properties of the total energy output of these sources, c) their time-evolution, d) their injection spectral 
properties of cosmic-ray $e^{\pm}$ and e) the propagation of cosmic rays through the ISM and the Heliosphere.
Then in section~\ref{sec:Data_and_Fitting}, we will discuss the data that we use and our fitting procedure.
In section~\ref{sec:Results}, we present our results. We show first our results from comparing to the 
observations above 15 GeV and then further discuss the lower energy analysis of measurements down to 5 GeV 
in the positron fraction and flux, where we notice a somewhat significant feature at $\simeq 12$ GeV.   
Finally, we give our conclusions and summary in section~\ref{sec:Conclusions}.
 
\section{Milky Way Pulsar Simulations}
\label{sec:Modeling}

Our pulsar simulations account for,
\begin{itemize}
\vspace{-0.14cm}
\item{the stochastic nature of the neutron stars' birth distribution in space and in time. We run simulations 
for different birth rates of neutron stars,} 
\vspace{-0.14cm}
\item{the initial conditions of the neutron stars in terms of their initial spin-down power distribution,}
\vspace{-0.14cm}
\item{the uncertainties on their time evolution, in terms of the braking index $\kappa$ and characteristic 
spin-down timescale $\tau_{0}$,}
\vspace{-0.14cm}
\item{the fraction of spin-down power that goes to cosmic-ray electrons and positrons released into the interstellar 
medium, and the injection spectrum these particles have,}
\vspace{-0.14cm}
\item{how the electrons/positrons propagate from the pulsars to us i.e. their propagation through the ISM 
and the Heliosphere.}
\end{itemize}

In this work we produce 7272 Milky Way pulsar simulations to account for the various combinations of 
parameters that we vary. Each of our simulations extends out to 4 kpc from the Sun and contains 
between $5\times10^{3}$ to $19\times10^{3}$ unique pulsars, depending on the exact assumption of 
the Milky Way pulsar birth rate. In the following we describe the specific assumptions that we test in 
our simulations. 

\subsection{The birth distribution of pulsars in space and time}
\label{subsec:PulsarSTdistribution}

From observations along the galactic plane we expect a pulsar birth rate of $1.4 \pm 0.2$ per century 
\cite{Lorimer:2006qs}. However, this rate is probably more uncertain as wider estimates have been 
made in~\cite{1999MNRAS.302..693D, Vranesevic:2003tp, FaucherGiguere:2005ny, Keane:2008jj}.
We take three basic choices for the pulsar birth rate of 0.6, 1 and 2 per century for the entire Milky Way.  
We note that more choices for the birth rate between the values of 0.6 and 2 pulsars per century would 
not change our basic results \footnote{Our simulations show a preference for a birth rate of 2 pulsars per 
century, but not in a manner that changes our conclusions.}.

In our simulation pulsars are stochastically generated based on a profile probability density function.
We follow the same spatial distribution as in earlier work of~\cite{Cholis:2017ccs}. Our 
spatial distribution relies on observations of Parkes multi-beam pulsar survey at 1.4
GHz \cite{Manchester:2001fp} and subsequent models in~\cite{FaucherGiguere:2005ny, 
Lorimer:2003qc,Lorimer:2006qs}. We assume azimuthal symmetry with respect to the galactic 
center and an exponentially decreasing pulsar density as we move away from the galactic disk, 
with a typical scale height of 50 pc. As most of the rotational energy of pulsars is emitted in their first 
$O(10)$ kyr, the associated natal kicks that are typically $O(10^2)$ km/s, can only result in a 
displacement of $O(1)$ pc. Such displacements represent a minor correction to the original birth
spacial distribution which we ignore.  

\subsection{The Pulsars' Initial Spin-Down Distribution Properties}
\label{subsec:PulsarInitialSpinDown}

Each pulsar due to asymmetries in the core-collapse of their progenitor star acquires an initial 
rotational energy. At the same time these are highly magnetized objects with initial B-field strengths 
at their poles of order $10^{12}$ G and up to $10^{15}$ G for magnetars. Typically the axis of 
rotation is not aligned to the axis of the magnetic field leading to energy losses, known as spin-down.
A pulsar's spin-down power $\dot{E}$ evolves with time and is modeled here as \cite{1973ApJ...186..249P},
\begin{equation}
\dot{E}(t) = \dot{E_{0}}  \bigg(1 + \frac{t}{\tau_{0}}  \bigg)^{-\frac{\kappa+1}{\kappa-1}}.
\label{eq:SpinDown}
\end{equation}
$\dot{E_{0}}$ is the original spin-down power of a given pulsar, $\tau_0$ is its characteristic 
spin-down timescale and $\kappa$ its braking index. For a given Milky Way pulsars 
simulation we assume that the pulsars' initial spin-down power is equal to $\dot{E_{0}}= 10^x$ 
ergs/s with $x = x_{\textrm{cutoff}} - y$. In each of our Milky Way simulations each pulsar has 
its unique $y$-value, i.e unique initial spin-down power.  Using the same parametrization 
of~\cite{Cholis:2017ccs}, the $y$-parameter follows a log-normal distribution, 
\begin{equation}
     f(y) = \frac{\textrm{Exp}\left\{ - \frac{[- \mu_{y} +
     ln(y)]^{2}}{2 \sigma_{y}^{2}} \right\}} {\sqrt{2 \pi} y
     \sigma_{y}}.
\label{eq:InitialSpinDownPower}
\end{equation}
The values for $x_{\textrm{cutoff}}$ and $\mu_{y}$ are constrained by radio observations  
and subsequent modeling of Myr old pulsars' periods \cite{FaucherGiguere:2005ny}; and 
by comparing to the ATNF pulsar catalog \cite{Manchester:2004bp, ATNFSite}. We set an
upper cutoff $x < x_{\textrm{max}} = 38.7$ indicative of the Crab pulsar's observed spin-down .
We take $\sigma_{y} = [0.25, 0.36, 0.5, 0.75]$ which allow varying degree of widths in those
distributions.  A $\sigma_{y} = 0$ would assume that all pulsars have an identical initial 
spin-down power. A distribution on the $y$-parameter should be expected both from 
the fact that there is a distribution on the magnitude of the initial B-fields at the poles and 
a distribution in the angle between the magnetic field axis and the axis of rotation.

\subsection{The Pulsars' Spin-Down Evolution}
\label{subsec:PulsarSpinDownEvolution}

\begin{figure}
\begin{centering}
\hspace{-0.6cm}
\includegraphics[width=3.72in,angle=0]{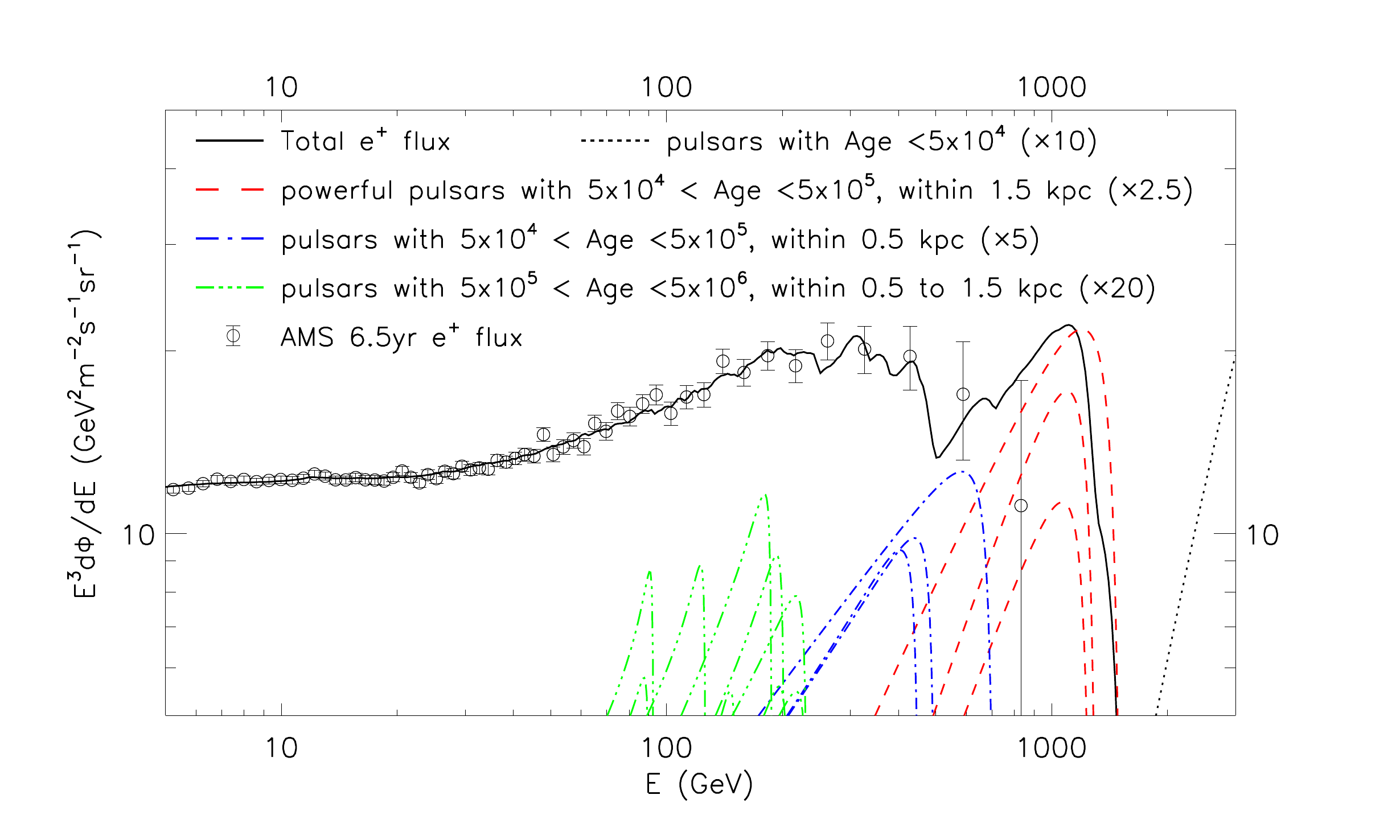}
\end{centering}
\vspace{-0.5cm}
\caption{The cosmic-ray positron flux form a Milky Way pulsars simulation. The solid black line includes 
the contribution of the cosmic-ray secondaries (from inelastic p-p, p-N and N-N collisions in the ISM). 
We highlight the contribution to the positron flux from individual pulsars that have ages from $O(10)$ kyr 
to 5 Myr and that are relatively close-by. The contribution from individual pulsars is enhanced from the 
original simulation to show their fluxes within the figure. The \textit{AMS-02} observation is shown in 
the data points.}
\vspace{-0.4cm}
\label{fig:PositronsFlux}
\end{figure}

As we said in the introduction we want to test the spin-down properties of pulsars relying on cosmic-ray 
observations. Detected cosmic-rays at the 5 GeV-5 TeV energy range can be used to probe the contribution
of pulsars with ages between $O(10^2)$ kyr and $O(10^2)$ Myr as we show in Figure~\ref{fig:PositronsFlux},
where we have highlighted the contribution of a few specific pulsars of given distance and age
ranges. Younger pulsars contribute at TeV energies. Middle-aged pulsars of age $10^{2}-10^{3}$ kyr 
contribute dominantly at $O(100)$ GeV and can give spectral features at these energies.
Older pulsars have suppressed overall fluxes but also more prominent peaks and subsequent cut-offs, due to 
cosmic-ray cooling, and might still be able to give minor spectral features at energies lower than 100 GeV. 

As we want to test if the braking index $\kappa$ is different for $O(10^{2})$-$O(10^{5})$ kyr 
pulsars compared to the much younger objects, we create Milky Way pulsars simulations where
all pulsar members have the same value of $\kappa$ and $\tau_{0}$. By creating such simulations
we can test if pulsars of older ages statistically prefer certain values of $\kappa$. If a pulsar's angle between
its axis of rotation and magnetic field axis evolves with time, that can be interpreted as an evolution 
of $\kappa$ \cite{Contopoulos:2005rs, Hamil:2016ylv}.  A relatively fast decay of the magnetic field's
amplitude at the poles can also lead to a changing braking index \cite{1988MNRAS.234P..57B, 
1992ApJ...395..250G, Tauris:2001cy, Marchant:2014ima, Gao:2017pxy, Igoshev:2021ewx}.  In addition, 
different equations of state can give for the same total mass different values of $\kappa$ 
\cite{Heiselberg:1998vh, Koliogiannis:2019rvh}. All these effects can result in the braking index evolving with 
time. Only a small number of pulsars exist with a reliably measured braking index, and all of them are 
very young \cite{Hamil:2015hqa, Archibald:2016hxz} (see however \cite{Parthasarathy:2020ksp}).  
Such young pulsars have negligible contribution to the observed cosmic-ray spectra and are not the 
focus of this study. 

In Table~\ref{tab:PulsarsSim}, we give all the spin-down power distribution properties 
for our Milky Way pulsars simulations. We test discrete values of $\kappa = [2.5, 2.75, 3.0, 3.25, 3.5]$, 
where $\tau_{0}$ vary from 0.6 kyr for $\kappa = 2.5$ to 30 kyr for $\kappa = 3.5$. These combination 
of values of $\tau_{0}$ and $\kappa$ are picked to be in agreement with surface magnetic fields as 
well as periods from \cite{FaucherGiguere:2005ny}. We also include different assumptions on the 
pulsars' distribution of the initial spin-down power. Table~\ref{tab:PulsarsSim}, is an expansion of 
earlier work in~\cite{Cholis:2017ccs}. We typically run 72 simulations per combination of $\tau_{0}$, 
$\kappa$, $x_{\textrm{cutoff}}$, $\mu_{y}$ and $\sigma_{y}$, but in some cases we run up to
108 (as in the 100-1H7) case. 

\begin{table}[t]
	\centering
	\small 
	\begin{tabular}{c | *{5}{c}} 

Sim no. 	
& $\tau_{0}$ (kyr) 	
& $\kappa$ 	
& $x_{\textrm{cutoff}}$ 	
& $\mu_{y}$ 	
& $\sigma_{y}$ 	
\\
\hline
		\hline
		100-1H7	
		& $6$ 	
		& $3$ 	
		& $38.8$ 	
		& $0.25$ 	
		& $0.5$ 	
		\\
		200-2H7	
		& $3.3$ 	
		& $3$ 	
		& $38.8$ 	
		& $0.25$ 	
		& $0.5$ 	
		\\
		300-3H7	
		& $10$ 	
		& $3$ 	
		& $38.8$ 	
		& $0.25$ 	
		& $0.5$ 	
		\\
		400-471	
		& $3.3$ 	
		& $3$ 	
		& $39$ 	
		& $0.1$ 	
		& $0.5$ 	
		\\
		500-571	
		& $1$ 	
		& $2.5$ 	
		& $38.8$ 	
		& $0.25$ 	
		& $0.5$ 	
		\\
		600-671
		& $20$ 	
		& $3.5$ 	
		& $39$ 	
		& $0.1$ 	
		& $0.5$ 	
		\\
		700-771
		& $0.7$ 	
		& $2.5$ 	
		& $38.8$ 	
		& $0.25$ 	
		& $0.5$ 	
		\\
		800-871
		& $20$ 	
		& $3.5$ 	
		& $39.1$ 	
		& $0.0$ 	
		& $0.25$ 		
		\\
		900-971
		& $0.6$ 	
		& $2.5$ 	
		& $39.0$ 	
		& $0.1$ 	
		& $0.25$ 	
		\\
		1000-1071
		& $6$ 	
		& $3$ 	
		& $39.0$ 	
		& $0.1$ 	
		& $0.25$ 	
		\\
		1100-1071
		& $6$ 	
		& $3$ 	
		& $38.7$ 	
		& $0.5$ 	
		& $0.75$ 	
		\\
		1200-1271 	
		& $30$ 	
		& $3.5$ 	
		& $38.8$ 	
		& $0.25$ 	
		& $0.5$ 	
		\\
		1300-1371
		& $0.85$ 	
		& $2.5$ 	
		& $38.5$ 	
		& $0.6$ 	
		& $0.75$ 	 	
		\\
		1400-1471
		& $18$ 	
		& $3.5$ 	
		& $39.0$ 	
		& $0.0$ 	
		& $0.75$ 		
		\\
		1500-1571 	
		& $10$ 	
		& $3$ 	
		& $38.7$ 	
		& $0.5$ 	
		& $0.75$ 	 	
		\\
		1600-1671
		& $4$ 	
		& $3$ 	
		& $39.0$ 	
		& $0.0$ 	
		& $0.36$ 		
		\\
		1700-1771 	
		& $1$ 	
		& $2.5$ 	
		& $38.7$ 	
		& $0.5$ 	
		& $0.75$ 	
		\\
		1800-1871
		& $9$ 	
		& $3$ 	
		& $38.2$ 	
		& $0.4$ 	
		& $0.36$ 	
		\\
		1900-1971
		& $0.8$ 	
		& $2.5$ 	
		& $38.2$ 	
		& $0.4$ 	
		& $0.36$ 	
		\\
		2000-2071
		& $0.6$ 	
		& $2.5$ 	
		& $38.2$ 	
		& $0.4$ 	
		& $0.36$ 	
		\\
		2100-2171
		& $30$ 	
		& $3.5$ 	
		& $38.2$ 	
		& $0.4$ 	
		& $0.36$ 	
		\\
		2200-2271
		& $7$ 	
		& $3$ 	
		& $39.0$ 	
		& $0.1$ 	
		& $0.75$ 	
		\\
		2300-23H7
		& $30$ 	
		& $3.5$ 	
		& $38.0$ 	
		& $0.5$ 	
		& $0.36$ 		
		\\
		2400-24H7	
		& $30$ 	
		& $3.5$ 	
		& $38.7$ 	
		& $0.5$ 	
		& $0.75$ 	 	
		\\
		2500-25H7
		& $6$ 	
		& $3$ 	
		& $38.9$ 	
		& $0.18$ 	
		& $0.36$ 	 	
		\\
		2600-26H7
		& $4.5$ 	
		& $3$ 	
		& $39.3$ 	
		& $0.0$ 	
		& $0.25$ 	 	
		\\
		2700-27H7
		& $9$ 	
		& $3$ 	
		& $38.5$ 	
		& $0.5$ 	
		& $0.25$ 	 	
		\\
		2800-28H7
		& $27$ 	
		& $3.5$ 	
		& $38.5$ 	
		& $0.3$ 	
		& $0.25$ 	
	         \\
		2900-29H7
		& $33$ 	
		& $3.5$ 	
		& $38.0$ 	
		& $0.5$ 	
		& $0.25$ 	
		\\
		3000-3071
		& $0.85$ 	
		& $2.5$ 	
		& $38.3$ 	
		& $0.5$ 	
		& $0.25$ 	
		\\
		3100-3171
		& $18$ 	
		& $3.25$ 	
		& $38.8$ 	
		& $0.25$ 	
		& $0.5$ 	
		\\
		3200-3271
		& $15$ 	
		& $3.25$ 	
		& $38.8$ 	
		& $0.25$ 	
		& $0.5$ 	
		\\
		3300-3371
		& $18$ 	
		& $3.25$ 	
		& $38.5$ 	
		& $0.4$ 	
		& $0.25$ 	
		\\
		3400-3471
		& $20$ 	
		& $3.25$ 	
		& $38.0$ 	
		& $0.4$ 	
		& $0.25$ 	
		\\
		3500-3571
		& $15$ 	
		& $3.25$ 	
		& $38.5$ 	
		& $0.4$ 	
		& $0.36$ 	
		\\
		3600-3671
		& $20$ 	
		& $3.25$ 	
		& $38.8$ 	
		& $0.5$ 	
		& $0.75$ 	
		\\
		3700-3771
		& $2$ 	
		& $2.75$ 	
		& $38.8$ 	
		& $0.25$ 	
		& $0.5$ 	
		\\
		3800-3871
		& $1.5$ 	
		& $2.75$ 	
		& $38.8$ 	
		& $0.25$ 	
		& $0.5$ 	
		\\
		3900-3871
		& $1.6$ 	
		& $2.75$ 	
		& $38.5$ 	
		& $0.5$ 	
		& $0.25$ 	
		\\
		4000-4071
		& $1.3$ 	
		& $2.75$ 	
		& $38.0$ 	
		& $0.4$ 	
		& $0.25$ 	
		\\
		4100-4171
		& $1.2$ 	
		& $2.75$ 	
		& $38.3$ 	
		& $0.4$ 	
		& $0.36$ 	
		\\
		4200-4271
		& $2.4$ 	
		& $2.75$ 	
		& $38.5$ 	
		& $0.5$ 	
		& $0.75$ 	
		\\

	\end{tabular}
	\caption{The Milky Way pulsars'-simulation spin-down power
        distribution and time evolution assumptions. We provide the 
        names here as a reference to our publicly available list of simulations\footnote{We 
        have made publicly available our Milky Way pulsars simulations through, 
        \texttt{https://zenodo.org/record/5659004\#.YYqnbi-ZN0s}.}.}
	\label{tab:PulsarsSim}
\end{table}

\subsection{Pulsars as Sources of Cosmic-Ray Electrons and Positrons}
\label{subsec:PulsarSources}

From microwave, X-ray and gamma-ray observations, we know that high-energy cosmic-ray electrons 
and positrons exist within a pulsar's magnetosphere. Moreover, electrons and positrons can be 
further accelerated in the termination shock of the pulsar wind nebula (PWN) and if there still is a supernova 
remnant shock (SNR), its termination shock just before entering the ISM. In addition observations from 
HAWC and Milagro show evidence for $O(10)$ TeV $\gamma$-rays towards Geminga and Monogem 
\cite{Abdo:2009ku, Abeysekara:2017hyn, Abeysekara:2017old}. These observations show that 10-100 TeV
cosmic-ray electrons and positrons exist at distances up to $\simeq 10$ pc from these middle aged pulsars.

In Ref.~\cite{Malyshev:2009tw}, it was shown that as a pulsar's characteristic spin-down timescale 
$\tau_{0}$ is $\sim 10^{4}$ yrs, an appreciable fraction ($\sim 1/2$) of the pulsar's initial rotational energy 
will be lost before the SNR shock front surrounding it stops being an efficient cosmic-ray accelerator. 
$\tau_{0}$ is also much smaller than the timescale the surrounding PWN stops accelerating cosmic-rays. 
As cosmic rays typically require $O(10^2)$ kyr to Myr to reach us from individual pulsars through 
diffusion, we can reliably approximate pulsars to instantaneously inject cosmic-ray electrons and 
positrons at the time of their birth (see Ref.~\cite{Malyshev:2009tw} for further details)
\footnote{An exception to that approximation would be the injection of cosmic-rays from a very young 
pulsar within $< 100$ pc. Such pulsars are very rare however, appearing in very few of our simulations. 
Moreover, as we show in Figure~\ref{fig:PositronsFlux}, their contribution would appear at very high 
energies, not observable by the satellite experiments we rely on.}.

The aim of this work is to constrain the averaged cosmic-ray injection spectral index $n$ from pulsars. 
For the injection spectra we assume,
\begin{equation}
\frac{dN}{dE} \propto  E^{-n} Exp \left\{ -\frac{E}{E_{\textrm{cut}}}\right\}.
\label{eq:InjSpect}
\end{equation}
We take $n$ to follow a uniform distribution within the range of $n \in \left[1.4,1.9\right]$ which we refer to 
as option "A" or two narrower ranges of $n \in \left[1.6,1.7\right]$, our option "B", or 
$n \in \left[1.3,1.5\right]$ our option "C". The upper cutoff $E_{\textrm{cut}}$ is taken to be  10 TeV.  
We find that its exact value does not affect our fitting results. This should be expected as the highest energy 
cosmic rays lose their energy faster. 
 
Furthermore, we model the fraction $\eta$  of rotational energy that ends in cosmic-rays released into the 
ISM and the relative variations of that fraction between pulsars. Following \cite{Cholis:2017ccs, Cholis:2018izy},  
we take a log-normal distribution for the $\eta$ parameter, 
\begin{equation}
g(\eta) = \frac{Exp \left\{ -\frac{ \left[ - \mu +ln(-1 + \eta) \right]^{2}}{2 \sigma^2}\right\}}{\sqrt{2 \pi} (\eta -1) \sigma},
\label{eq:eta}
\end{equation}
and take three different choices for ($\mu$, $\sigma$) to be (0.32, 0.12) (option "A"), (0.64,0.23) (option "B") 
and (-0.38, 0.16) (option "C"). These give square root variances of 0.169, 0.454 and 0.112 respectively. 
In our Milky Way pulsars simulations before fitting to the data we also pick specific values for $\mu$ that affect 
the mean efficiency of these pulsars, 
$\bar{\eta} = 1 + \textrm{Exp}\left\{ \mu + \frac{\sigma^2}{2} \right\}$. These  are $\bar{\eta} = 4\times 10^{-3}$ 
(for option "A"), $10^{-3}$ (for option "B") and $2\times 10^{-2}$  (for option "C"). However, we fit each Milky 
Way pulsars simulation to the data and thus the $\bar{\eta}$ is reset by the data. 

\begin{figure}
\begin{centering}
\hspace{-0.6cm}
\includegraphics[width=3.72in,angle=0]{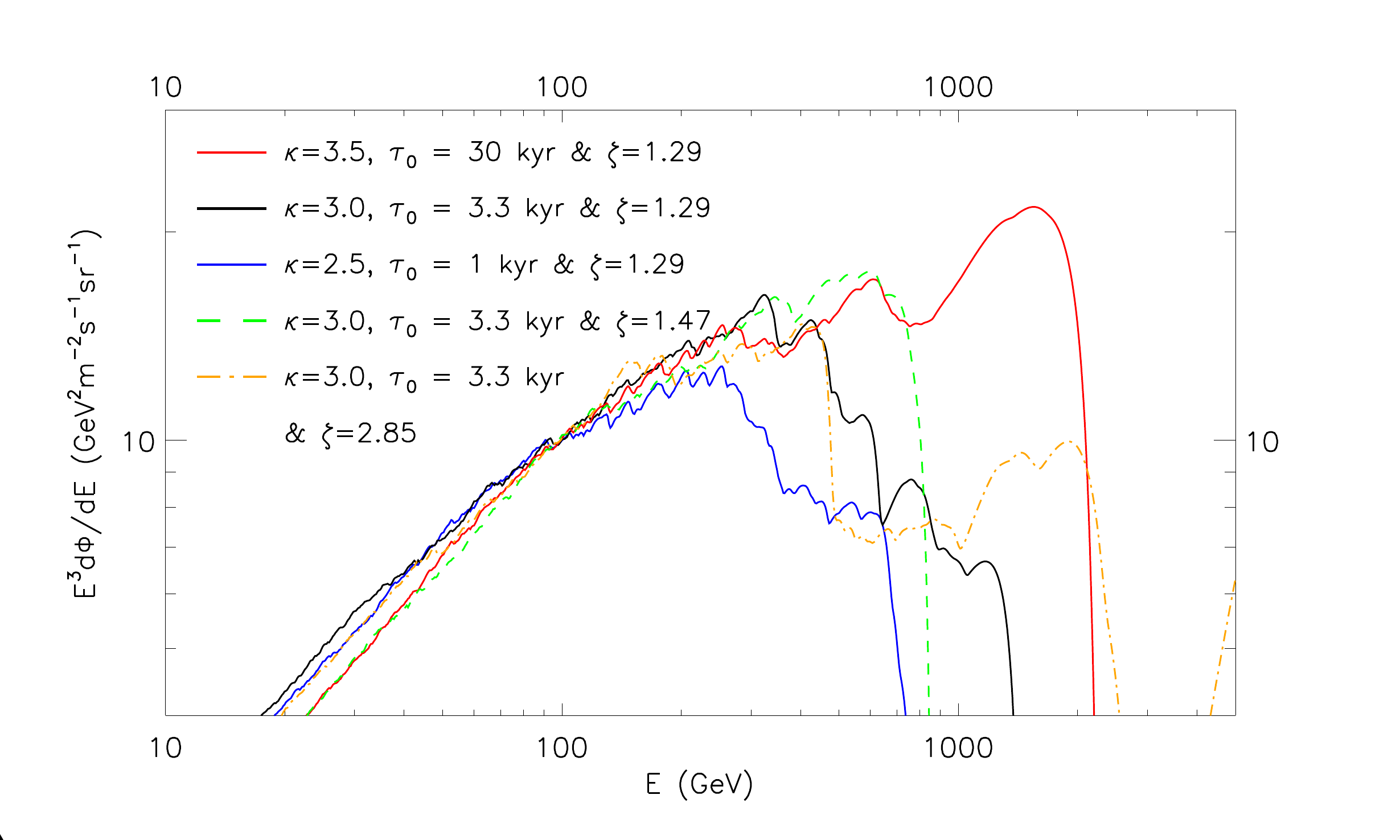}
\end{centering}
\vspace{-0.5cm}
\caption{The positron flux for different choices of spin-down evolution (red vs black vs blue) for a fixed assumption 
on $\zeta = 10^{\sqrt\textrm{variance}}$. We also compare different choices for $\zeta$ by keeping the spin-down 
assumptions fixed (black solid vs  green dashed vs orange dot-dashed). See text for details.}
\vspace{-0.4cm}
\label{fig:Kappa_Tau0_Zeta}
\end{figure}

In Figure~\ref{fig:Kappa_Tau0_Zeta}, we show five simulations of Milky Way pulsars to highlight the impact of 
our assumptions on the barking index $\kappa$, the spin-down timescale $\tau_{0}$ and the standard deviation 
of the fraction of energy going to cosmic-ray electrons and positrons variance, where $10^{\sqrt\textrm{variance}} = \zeta$. 
We simulate $8.6 \times 10^{3}$ individual pulsars, whose locations and ages are fixed. The youngest of these 
pulsars is 175 yr and the oldest 10 Myr, and all are with 4 kpc from us. We change the spin-down evolution of 
those pulsars by assuming different values of $\kappa$ and $\tau_{0}$ relevant in Eq.~\ref{eq:SpinDown}, 
taking a fixed value of $\zeta$ (red vs black vs blue solid lines). We normalize all simulations in Figure 
\ref{fig:Kappa_Tau0_Zeta} to get the same positron flux at 100 GeV. This is done to showcase the impact of 
these assumptions in our analysis. As we will describe in section~\ref{sec:Data_and_Fitting}, we fit the simulated
fluxes to the \textit{AMS-02} observed flux. Once fitting to the positron flux, a higher value of $\kappa$, would 
suggest that most flux from pulsars and the associated features are at higher energies. When pulsars have a larger 
$\kappa$ they release more slowly their energy. As a result they can remain strong sources of high-energy cosmic 
rays for a longer amount of time. In turn their combination gives enhanced fluxes at the highest energies. 

In Figure~\ref{fig:Kappa_Tau0_Zeta}, we also show the impact of varying the assumptions on $\sigma$, by fixing 
the spin-down evolution to $\kappa = 3.0$ and $\tau_{0} = 3.3$ kyr. Larger values of $\sigma$ in Eq.~\ref{eq:eta}, 
result in some pulsars depositing a great fraction of their energy in electrons and positrons. Such pulsars are 
responsible for prominent features in the positron flux (and the positron fraction). A similar effect exists if instead 
pulsars have a large variance in their initial spin-down power, i.e. the value of $\sigma_{y}$ of 
Eq.~\ref{eq:InitialSpinDownPower} is chosen to be high.  

\subsection{Cosmic-Ray Propagation through the ISM and heliosphere}
\label{subsec:CRprop}

Cosmic-ray electrons and positrons propagate in the interstellar medium via diffusion. How fast cosmic rays 
diffuse depends on the galactic magnetic field's amplitude and structure, and their energy. We assume isotropic
and homogeneous diffusion that can be described by a rigidity ($R$)-dependent diffusion coefficient,
\begin{equation}
D(R) = D_{0} \left( \frac{R}{1 \, GV} \right)^{\delta}.
\label{eq:diffusion}
\end{equation}
$D_{0}$ is the relevant normalization set at 1 GV, regulated by the overall strength of magnetic fields in the 
Milky Way, while the diffusion index $\delta$ is defined by the spectrum of interstellar turbulence. $\delta = 0.33$, 
is for the case of Kolmogorov turbulence \cite{1941DoSSR..30..301K}, while $\delta = 0.5$ is for Kraichnan 
two-dimensional turbulence \cite{1967PhFl...10.1417K}. Systematic study of cosmic ray observations suggest 
that the diffusion index $\delta$ is within that range of values (see e.g. \cite{Trotta:2010mx, Pato:2010ih}); which 
we use in our simulations. We model the diffusion taking place within a cylinder centered at the galactic center, 
of radius 20 kpc and extending to a height of $ \pm z_{L}$ away from the galactic disk (the disk is at $z=0$). 
Beyond those limits cosmic rays will escape the galaxy. 

Furthermore, cosmic-ray electrons and positrons at the 10 GeV to 10 TeV range lose rapidly energy through 
inverse Compton scattering and synchrotron radiation with a rate that scales as,
\begin{equation}
\frac{dE}{dt} = -b \left( \frac{E}{1 \, GeV} \right)^{2}.
\label{eq:Eloss}
\end{equation}
This makes the highest energy cosmic-ray electrons and positrons lose more rapidly their energy and causes
a "pile-up" before the cut-off, in the electron/positron fluxes from individual pulsars that is seen in Figure 
~\ref{fig:PositronsFlux}. The value of $b$, set at 1 GeV, is directly proportional to the energy density in the 
galactic magnetic field and the energy density in the CMB and interstellar radiation field photons. We note that 
at the highest energies the inverse Compton cross-section is not the Thomson cross-section assumed in Eq.
~\ref{eq:Eloss}, but instead the Klein-Nishina one \cite{1929ZPhy...52..853K, 1970RvMP...42..237B}. In our
pulsars simulations we ignore the Klein-Nishina corrections, as we use a wide range of uncertainty on the $b$ 
parameter, set to be within $3\times 10^{-6}$ and $8 \times 10^{-6}$ GeV$^{-1}$ kyrs$^{-1}$. We also ignore 
Bremsstrahlung emission losses that cause an energy-loss rate $\propto E$, that become important only at 
GeV energies.

Cosmic rays also experience diffusive reacceleration \cite{1994ApJ...431..705S} and are affected by local 
convective winds. To account for ISM diffusion uncertainties we use four distinctive models, defined by the 
letters A, C, E, F. These models are in agreement with \textit{AMS-02} observations of the cosmic-ray proton, 
helium, carbon, oxygen fluxes and the beryllium-to-carbon, boron-to-carbon and oxygen-to-carbon ratios 
\cite{Cholis:2021inprep}. Each one of these models has three variants to account for uncertainties in the 
energy losses, i.e the $b$-parameter, 
denoted by a second character (1-3). A value of  $b = 5.05 \times 10^{-6}$GeV$^{-1}$kyrs$^{-1}$ (for models 
A1, C1, E1, and F1), is in agreement with evaluations of the local magnetic and interstellar radiation field 
\cite{galprop, GALPROPSite, Porter:2017vaa}, while the choices of $2.97  \times 10^{-6}$ and $8.02 \times 10^{-6}$ 
GeV$^{-1}$ kyrs$^{-1}$ represent the relevant uncertainties. All these parameters are described in 
Table~\ref{tab:ISMBack}.

\begin{table}[t]
    \begin{tabular}{ccccc}
         \hline
           Model &  $z_{L}$ (kpc) & $b$ ($\times 10^{-6}$GeV$^{-1}$kyrs$^{-1}$) & $D_{0}$ (pc$^2$/kyr) & $\delta$\\
            \hline \hline
            A1 &  5.7 & 5.05 & 140.2 & 0.33 \\
            A2 &  5.7 & 8.02 & 140.2 & 0.33 \\
            A3 &  5.7 & 2.97 & 140.2 & 0.33 \\      
            C1 &  5.5 & 5.05 & 92.1 & 0.40 \\
            C2 &  5.5 & 8.02 & 92.1 & 0.40 \\
            C3 &  5.5 & 2.97 & 92.1 & 0.40 \\      
            E1 &  6.0 & 5.05 & 51.3 & 0.50 \\
            E2 &  6.0 & 8.02 & 51.3 & 0.50 \\
            E3 &  6.0 & 2.97 & 51.3 & 0.50 \\
            F1 &  3.0 & 5.05 & 33.7 & 0.43 \\
            F2 &  3.0 & 8.02 & 33.7 & 0.43 \\
            F3 &  3.0 & 2.97 & 33.7 & 0.43 \\
        \hline \hline 
        \end{tabular}
\caption{The ISM parameters that describe the propagations assumptions of cosmic rays in the Milky Way.} 
\label{tab:ISMBack}
\end{table}

In comparing the ISM model predictions to the \textit{AMS-02} cosmic-ray nuclei measurements we used
\path{GALPROP} v54 \cite{galprop, GALPROPSite} where we have included convection, reacceleration and
Bremsstrahlung energy-losses. \path{GALPROP} gives us a prediction for the primary cosmic-ray electrons 
(from SNRs) and the secondary electrons and positrons from $p-p$, $p-N$ and $N-N$ inelastic collisions 
taking place at the ISM gas. However, for the cosmic ray electrons and positrons from the individual pulsars 
the code that we use ignores ISM convection, diffusive reacceleration and Bremsstrahlung energy 
losses \cite{Malyshev:2009tw, Cholis:2017ccs, Cholis:2018izy}. In the energies of interest the timescale 
for these effects are significantly larger than the diffusion and energy losses of Eq.~\ref{eq:Eloss} timescales.
This allows us to place within 4 kpc from the location of the Sun up to $1.9 \times 10^{4}$ pulsars in unique
positions and of unique age created in the last 10 Myr. 
We take the primary electron fluxes and secondary electron and positron fluxes from \path{GALPROP} and 
combine them with the pulsars' electron and positron fluxes from our Milky Way pulsars simulations.

\begin{figure}
\begin{centering}
\hspace{-0.6cm}
\includegraphics[width=3.72in,angle=0]{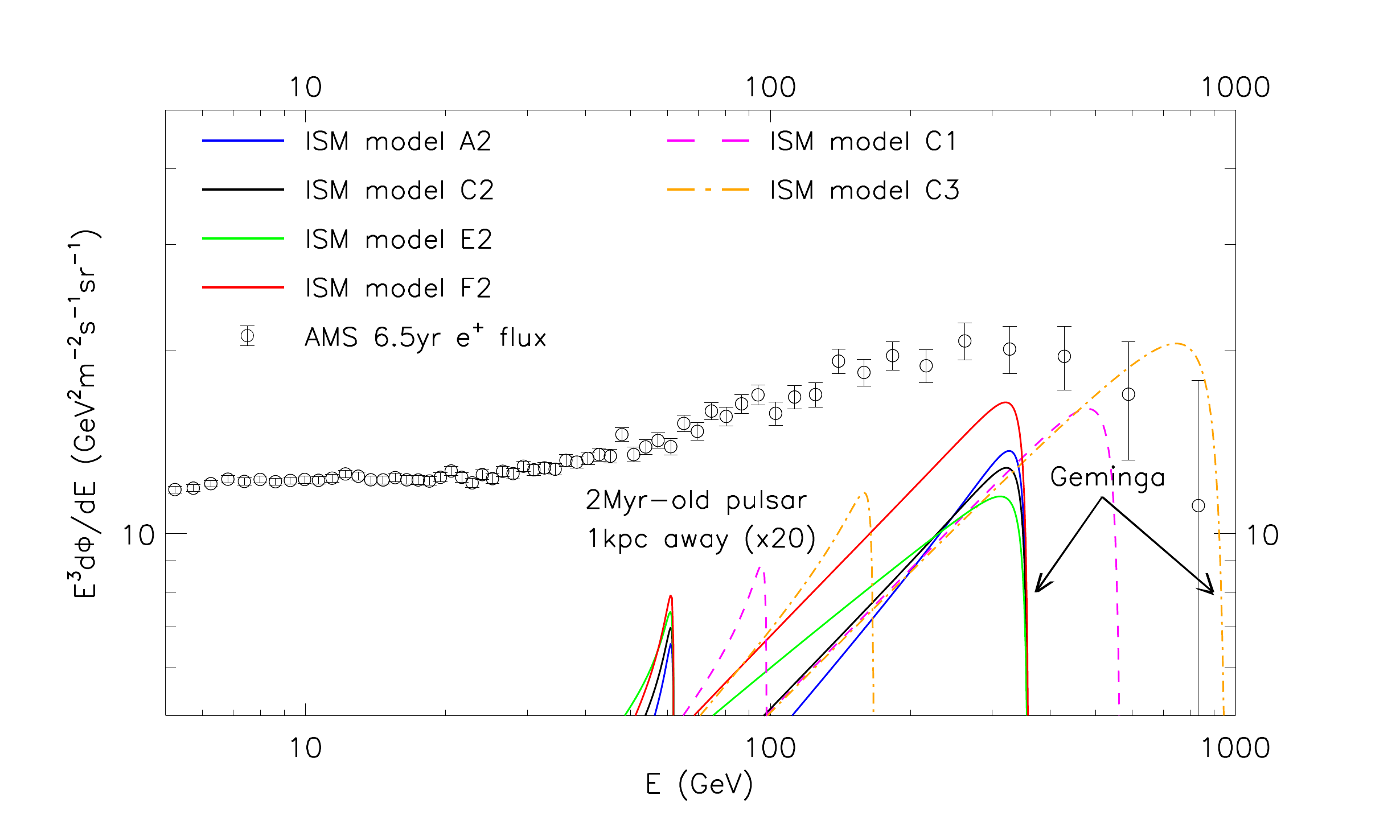}
\end{centering}
\vspace{-0.5cm}
\caption{The impact of different ISM assumptions on the positron flux from two individual pulsars. Solid lines 
(blue, black, green and red) show the impact of different diffusion assumptions. The magenta dashed and 
orange dot-dashed are to be compared to the black solid line that is made under the same diffusion assumptions.}
\vspace{-0.4cm}
\label{fig:ISM_Prop}
\end{figure}

In Figure~\ref{fig:ISM_Prop}, we show how the positron flux from individual pulsars is affected by different 
ISM propagation conditions. We model two individual pulsars, Geminga that is taken to be 0.25 kpc away 
and $3.42 \times 10^{5}$ yr in agreement with observations \cite{ATNFSite}, and a second pulsar 1.0 
kpc away and $2.0 \times 10^{6}$ yr old. Both pulsars are taken to have an initial spin-down power of 
$\dot{E}_{0} = 1.1 \times 10^{38}$ erg/s, braking index $\kappa = 3.0$ and $\tau_{0} = 6.0$ kyr, that 
for Geminga would give the currently observed spin-down power of $3.2 \times 10^{34}$ erg/s. The 
cosmic-ray injection index of Eq.~\ref{eq:InjSpect} is taken to be $n=1.65$ for both. 

Keeping energy losses fixed and changing between A2, C2 and E2 we note the difference in the positron 
flux's power-law spectrum for energies lower than the sharp cooling cut-off. That is simply the affected by
the diffusion index $\delta$. For model F2 the flux is significantly larger as the diffusion normalization $D_{0}$ 
is the smallest to account for its small scale height of $z_{L} = 3$ kpc. Cosmic-ray electrons/positrons 
from close-by pulsars stay longer close of their source for smaller values of $D_{0}$. That is most 
evident in Figure~\ref{fig:ISM_Prop} with Geminga. The energy losses set the value of the cooling cut-off. 

The observed cosmic-ray spectra by \textit{AMS-02} are affected by solar modulation. Cosmic rays have 
to travel through the Heliosphere before being recorded. During their propagation through the Heliosphere,
cosmic rays diffuse through a fast evolving anisotropic magnetic field, experience drift effects and adiabatic 
losses. As a result their energy gets statistically shifted to lower values, described by the solar modulation potential 
$\Phi$ \cite{1968ApJ...154.1011G}. We use here the time-, charge- and rigidity-dependent formula for the 
solar modulation potential from \cite{Cholis:2015gna}, that has recently been further constrained in 
\cite{Cholis:2020tpi}. Our model for solar modulation requires as inputs the value of the total B-field of 
the Solar Wind at 1 AU, which we take from \textit{ACE} \cite{ACESite}, the tilt angle of the heliospheric 
current sheet that is modeled in Wilcox Solar Observatory \cite{WSOSite} and the polarity of the Heliospheric 
magnetic field. In turn it gives us a value for the solar modulation potential that the kinetic energy of a 
particle of mass $m$, rigidity $R$ and charge $q$ was shifted by at a given Bartels' Rotation number. 
Our solar modulation model has two free-parameters $\phi_{0}$ and $ \phi_{1}$ that we marginalize over 
within a range of values most recently constrained in \cite{Cholis:2020tpi}. For further details see 
\cite{Cholis:2015gna, Cholis:2020tpi}. 

\section{Cosmic ray data and fitting procedure}
\label{sec:Data_and_Fitting}

In this section we describe the observational data that we use to test our pulsars population models and the 
specifics of the fitting procedure that we follow. We account for uncertainties in the overall normalization of the 
pulsars' contribution, and for spectral and normalization uncertainties of other components in the electron and
positron cosmic-ray spectra. 

\subsection{Observations of cosmic-ray electrons and positrons}
\label{sec:Data}

We use the publicly available \textit{AMS-02} positron flux, the positron fraction and the $e^{+} + e^{-}$ flux from
\cite{AMS:2021nhj, AMS:2019iwo, AMS:2019rhg}. Specifically for the $e^{+} + e^{-}$ flux we rely on \cite{AMS:2021nhj} 
instead of \cite{AMS:2019iwo}.  The analysis of \cite{AMS:2021nhj} avoids charge sign identification and therefore 
results in a higher efficiency. For the $e^{+} + e^{-}$ flux measurements we start at 10 GeV to avoid possible low 
energy systematics that may exist in comparing the \textit{AMS-02} measurement to that of \textit{DAMPE} 
\cite{DAMPE:2017fbg} or \textit{CALET} \cite{Adriani:2018ktz}.
For the positron flux and the positron fraction, we used the measurements at energies of 5 GeV or higher. At lower 
energies the pulsars' contribution is expected to be subdominant and mostly affected by pulsars several kpc away 
that we do not model.
Moreover, the lower-energy $e^{\pm}$ spectra are strongly affected by solar modulation that we account for, but also other 
propagation uncertainties as those of diffusive re-acceleration and cosmic-ray convection in the ISM that we 
set to be present but do not further marginalize over.

In addition, we perform fits ignoring the measurements below 15 GeV. At $\sim 12$ GeV there is a bump at the positron 
fraction that our simulations find it difficultly to fit. We will come back to the matter of possible interpretations to this
bump in section~\ref{sec:Conclusions}.
Finally, we use the $e^{+} + e^{-}$ flux measurements from \textit{DAMPE} \cite{DAMPE:2017fbg} and \textit{CALET} 
\cite{Adriani:2018ktz}. The used datasets with their respective data acquisition era and the corresponding Bartels' 
Rotation (BR) Numbers are presented in Table \ref{tab:datasets}.

\begin{table}[t]
    \begin{tabular}{c||c|c|c}
         \hline
           Dataset & Acquisition Era & BR \# & Ref.\\
            \hline \hline
            \textit{AMS-02} $e^+$ & 5/2011 - 11/2017 & 2426 - 2514 & \cite{AMS:2019rhg} \\
            \textit{AMS-02} $e^+/(e^-+e^+)$ & 5/2011 - 11/2017 & 2426 - 2514 & \cite{AMS:2019iwo} \\
            \textit{AMS-02} $e^-+e^+$ & 5/2011 - 11/2017 & 2426 - 2514 & \cite{AMS:2021nhj} \\
            \textit{DAMPE} $e^-+e^+$ & 12/2015 - 6/2017 & 2488 - 2508 & \cite{DAMPE:2017fbg} \\
            \textit{CALET} $e^-+e^+$ & 10/2015 - 11/2017 & 2486 - 2515 & \cite{Adriani:2018ktz} \\
            \hline \hline 
        \end{tabular}
\caption{The cosmic-ray measurements from \textit{AMS-02}, \textit{DAMPE} and \textit{CALET} used in this analysis.} 
\label{tab:datasets}
\end{table}

\subsection{Fitting}
\label{sec:Fitting}

We allow for up to seven free parameters to be optimized in our simulations. 
Two solar modulation parameters $\phi_0$ and $\phi_1$, three normalization factors  $a$, $b$ and $c$ 
for the primary cosmic-ray $e^-$ flux, secondary cosmic-ray $e^{\pm}$ fluxes and total pulsar $e^{\pm}$ fluxes, and 
two spectral indices $d_1$ and $d_2$ responsible for hardening or softening the primary $e^-$ and 
secondary $e^{\pm}$ spectra by multiplying them with $(E/1\GeV)^{d_1}$ and $(E/100\GeV)^{d_2}$ 
respectively. We remind to the reader that the primary and secondary fluxes are already modeled to include 
specific energy losses, diffusion, diffusive re-acceleration, convection and ISM gas distribution assumptions. 
They also originate from a distribution model for all primary cosmic-ray sources. The additional normalizations
($a$, $b$) and spectral freedoms ($d_{1}$, $d_{2}$) are to account for uncertainties in the overall efficiency 
and number of the primary cosmic-ray sources, the total ISM gas density, the exact injection spectra of 
primary $e^{-}$ spectra and cosmic-ray nuclei spectra that through their collisions give the secondary 
$e^{\pm}$.   

We fit each produced simulation to each dataset via a $\chi^2$ minimization. While the \textit{AMS-02} 
measurements were acquired during the same era, we do not fit all the \textit{AMS-02} datasets 
simultaneously as they originate using different type of analysis cuts. Similarly, we do not try to fit 
simultaneously all $e^-+e^+$ measurements from the three different experiments, as some are 
in statistical tension with each other. 

We first fit each simulation to the \textit{AMS-02} positron flux. Then we focus on the realizations that 
can fit to the positron flux data within 2$\sigma$, 3$\sigma$ or 5$\sigma$ from an expectation of $\chi^2$ 
of 1 for each degree of freedom (d.o.f.). For our fits of the the positron spectrum above 5 GeV there are 59 
data points being fitted with five parameters. Thus the 3$\sigma$ and 5$\sigma$ limits that we use translate 
to a $\chi^2$/d.o.f. of 1.337 and 1.683 respectively. Instead, for the positron spectrum in energies $E \ge 15$ 
GeV there are 44 data points. The relevant 2$\sigma$ and 3$\sigma$ limits we present translate to $\chi^2$/d.o.f. of 
1.290 and 1.467.
In those fits we take $\phi_{0}$ and $\phi_{1}$ to be within $[0.1,0.6]$ GV and $[0,2]$ GV respectively.
Our range for the normalization $b$ is $[0.8,2]$. The parameter $c$ is only given an upper bound such 
that $\bar{\eta}\ c\leq 0.5$ for each realization, while the parameter $d_{2}$ is within $[-0.1,0.1]$.

We repeat the same fitting procedure with the \textit{AMS-02} positron fraction and $e^-+e^+$ flux; where
for those we make use of all seven free parameters. We focus only on the simulations that are within
the 2$\sigma$, 3$\sigma$ or 5$\sigma$ positron flux limits. For each of those simulations we take the 
best fit values of $b$, $c$ that we got from the relevant positron flux fit and allow for up to a 
50\% additional variation \footnote{We always retain the hard limit of 
$\bar{\eta}\ c\leq 0.5$ that originates from equipartition of spin-down power to cosmic rays and B-field.}. 
Parameters $\phi_{0}$, $\phi_{1}$ are taken within the same ranges reported above as we consider 
these to be quite wide, while $d_{2}$ is fixed at its best fit value from the positron flux. Finally, the two newly 
introduced parameters $a$ and $d_{1}$ relating to the primary $e^{-}$ component take values within 
the range of $[0.6,1.2]$ and $[-0.2,0.5]$ respectively. 

We report the simulations that can fit each of the  \textit{AMS-02} data, i) within 2$\sigma$ and 3$\sigma$ 
for energies of 
$E> 15$ and ii) within 3$\sigma$ and $5\sigma$ for $E \ge 5$ GeV for the positron flux and fraction and  $E> 10$
for the total $e^-+e^+$ flux.
We then compare the retained simulations from our $E> 15$ fits to the 
\textit{DAMPE} and \textit{CALET} data. The \textit{DAMPE} and \textit{CALET} fits are performed in the 
same manner as the \textit{AMS-02} $e^{+}+e^{-}$ flux.  

When we fit the positron fraction in each energy bin we use counts instead of fluxes as \textit{AMS-02} uses 
binned data. At the highest energies some of our simulations may suggest the presence of multiple features 
within the same energy bin. Thus, we integrate the differential flux $d\Phi^{e^{\pm}}/dE$ over the energy range 
of each bin, and set a number of positron and electron events $N^{e^{\pm}}$ as,
\begin{equation}
\label{eq:flux_to_counts}
N^{e^{\pm}}=\int_{E_{\textrm{min}}}^{E_{\textrm{max}}}\dfrac{d\Phi^{e^{\pm}}}{dE} dE,
\end{equation}
where $E_{\textrm{min}}$, $E_{\textrm{max}}$ are the bounds of the bin. We note that $N^{e^{\pm}}$ is not 
the real event count in each bin as we don't know the exact exposure of the \textit{AMS-02} detector. For any 
given energy bin assuming the exposure is roughly constant, ignoring it, is not an issue as the exposure cancels 
in taking the positron fraction ratio.

The positron fraction becomes,
\begin{equation}
\label{eq:pf_counts}
\dfrac{b N_{e^+}^{sec}+c N_{e^+}^{pul}}{a N_{e^-}^{pri}+b N_{e^-}^{sec}+c N_{e^-}^{pul}+
b N_{e^+}^{sec}+c N_{e^+}^{pul}}.
\end{equation}
The indices ``pri'', ``sec'', ``pul'' refer to the primary, secondary and pulsar fluxes respectively. We note that the 
factors $(E/1\GeV)^{d_1}$ and $(E/100\GeV)^{d_2}$ have been absorbed into their respective counts. 
For the positron flux or the $e^{+}+e^{-}$ flux fits using "counts" is not possible as we do not know the 
exact exposures. Therefore we remain on working with fluxes. 

Our minimization procedure uses \path{SciPy}'s \cite{2020SciPy-NMeth} \path{least_squares} routine 
from the \path{optimize} module to solve our non-linear least squares problem. We also tried \path{iMinuit} 
\cite{iminuit} but found out that the \path{least_squares} achieves a good minimization much faster
\footnote{We noticed that the starting value for the parameter $d_{1}$ can cause the minimization 
in the positron fraction to fall into local minima. Therefore we minimize each simulation several times, 
starting from different values for $d_{1}$ in its allowed range. In each minimization all the other parameters' 
starting values are chosen randomly within their allowed ranges as we found that their exact starting value 
doesn't cause any issues.}.

\section{Results}
\label{sec:Results}

The first step in testing every Milky Way pulsars simulation is to fit its predicted positron flux in 
combination with a secondary positron flux component to the \textit{AMS-02} positron flux measurement,
as shown for one model in Figure~\ref{fig:PositronsFlux}. The second step is to test that simulation 
against the positron fraction measurement and the $ e^{+} + e^{-}$  measurement (see Section~\ref{sec:Fitting} for details). 
Our secondary and primary flux components are evaluated for the same diffusion and energy losses assumptions 
as the electrons and positrons fluxes originating from pulsars. 

\begin{figure}
\begin{centering}
\hspace{-0.6cm}
\includegraphics[width=3.72in,angle=0]{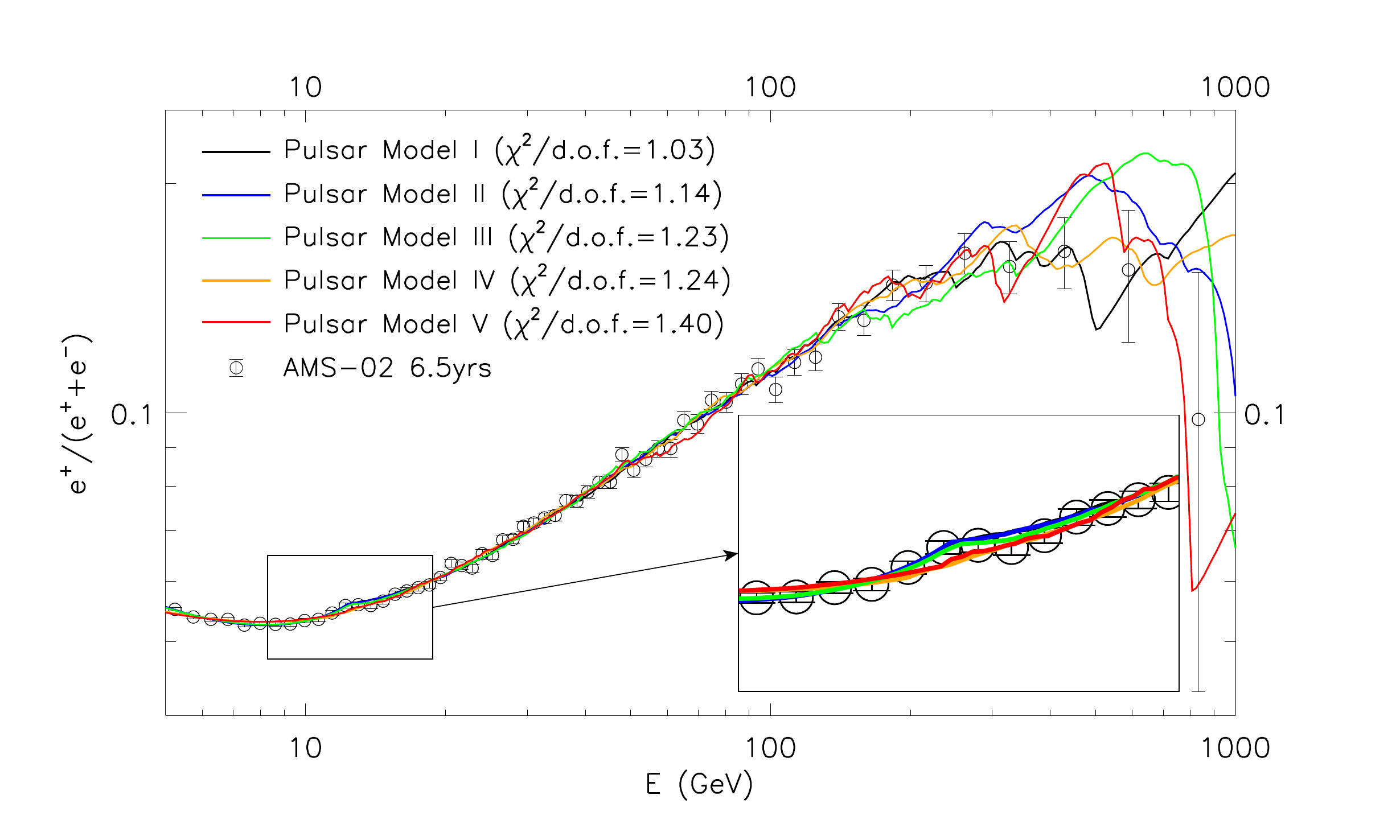}
\end{centering}
\vspace{-0.5cm}
\caption{The predicted positron fraction spectra for five different pulsar realizations along with the 
\textit{AMS-02} data. The $\chi^2$/d.o.f. of these pulsar models are also shown, evaluated at 
energies greater than 5 GeV. The zoomed part shows the spectrum around the bump that is 
centered at $\simeq 12$ GeV.}
\vspace{-0.4cm}
\label{fig:PositronFraction}
\end{figure}

In Figure~\ref{fig:PositronFraction}, we show the positron fraction spectra from different pulsars simulation 
assumptions. There is a clear feature that peaks around 12 GeV.  A similar feature exists also in the positron 
flux. In presenting our results we break our discussion into two subsections. The first one (\ref{subsec:Results15GeV}) 
is focused on fitting only energies above 15 GeV avoiding the impact of a such a feature. Instead in 
Section~\ref{subsec:Results5GeV}, we include the energies of 5 to 15 GeV in our fits.

Pulsar model I of Figure~\ref{fig:PositronFraction}, represents the best fit in the positron 
fraction from energies 5 GeV and above. This same model gives a good fit at $E>15$ GeV 
as well. Our simulations include enough 
physical variations that the predicted positron fraction spectrum, which increases from 
7 to $\sim 300 \GeV$ can either keep rising at higher energies (model I), drop (model II,III,V) 
or flatten out (model IV). We also note that our allowed pulsars simulations can have a 
noticeable amount of features in them, something that can be searched for independently
(see \cite{Cholis:2017ccs} for a detailed discussion).

\begin{figure}
\begin{centering}
\hspace{-0.6cm}
\includegraphics[width=3.72in,angle=0]{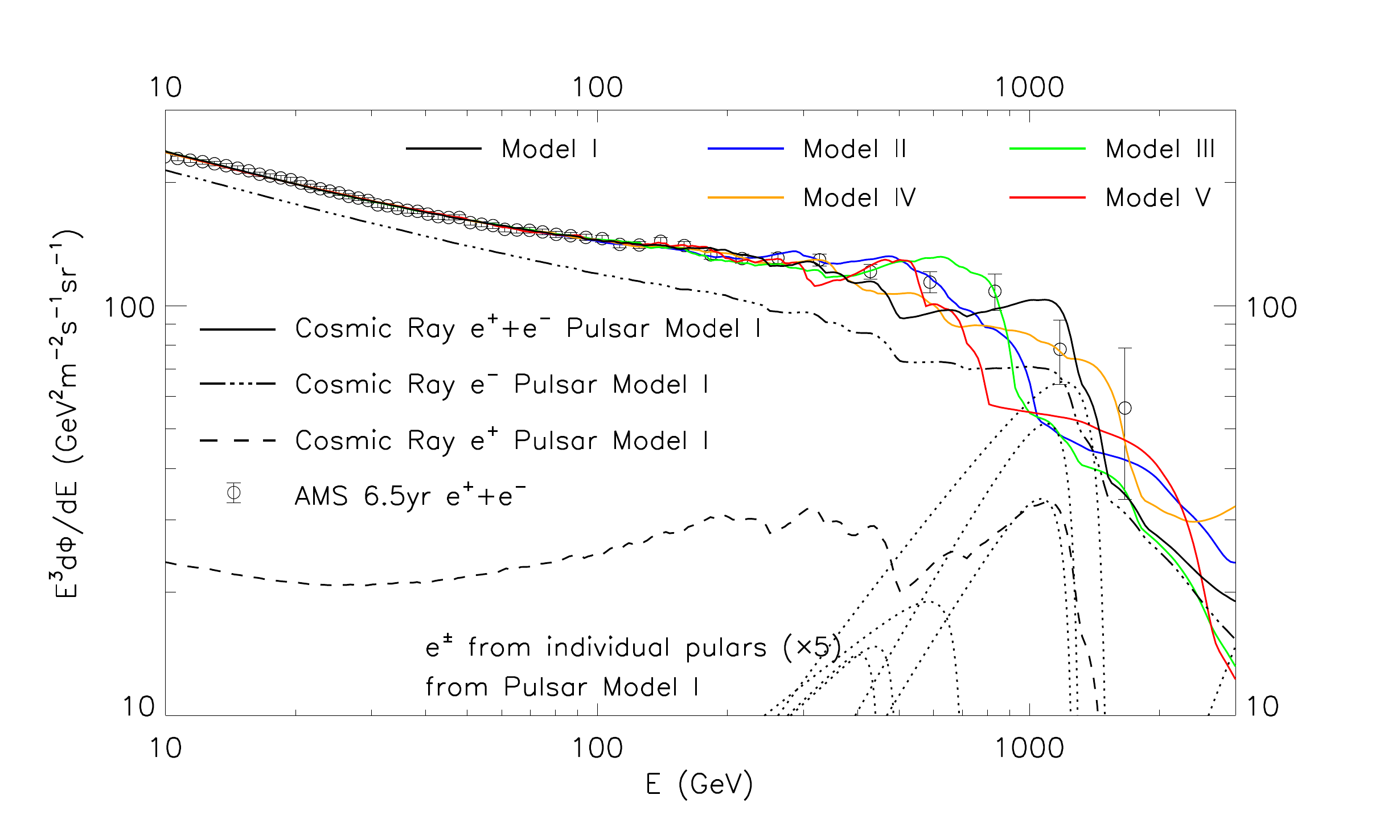}
\end{centering}
\vspace{-0.5cm}
\caption{The predicted $e^{+} + e^{-}$ spectrum for the same five pulsar simulations of 
Figure~\ref{fig:PositronFraction}, along with the \textit{AMS-02} observation. For model I, 
we also show the individual electron and positron spectra and fluxes from individual pulsars.}
\vspace{-0.4cm}
\label{fig:TotalLeptonFlux}
\end{figure}

In Figure~\ref{fig:TotalLeptonFlux}, we depict the five $e^{+}+e^{-}$ spectra for the same pulsar 
models (I through V). For model I we plot the $e^-$ and $e^+$ fluxes separately and the 
fluxes from some individual pulsars. These simulations can fit the \textit{AMS-02} positron 
and the $e^{+}+e^{-}$ fluxes well and have $\chi^2/\mathrm{d.o.f.}<1$, with the exception of 
model V that has $\chi^2/\mathrm{d.o.f.}=1.51$ in its $e^{+}+e^{-}$ fit. Similarly to what was
shown in Figure~\ref{fig:PositronsFlux} for the positrons, individual pulsars can give features 
in the higher energies of the electron spectrum and possibly explain features in the combined 
$e^{+}+e^{-}$ spectrum. 

\subsection{Results for $E>15\ \textrm{GeV}$ fits}
\label{subsec:Results15GeV}

\subsubsection{Using only \textit{AMS-02} data}

Of the 7272 astrophysical realizations, 2105 can fit the \textit{AMS-02} positron spectrum within 
$3\sigma$ from an expectation of $\chi^2$ of 1 for each degree of freedom. Of these 2105 
realizations,  567 (1095) can also fit within $2\sigma \, (3\sigma)$ the \textit{AMS-02} positron 
fraction spectrum and $e^{+}+e^{-}$ spectrum.

\begin{figure}
\begin{centering}
\hspace{-0.6cm}
\includegraphics[width=3.72in,angle=0]{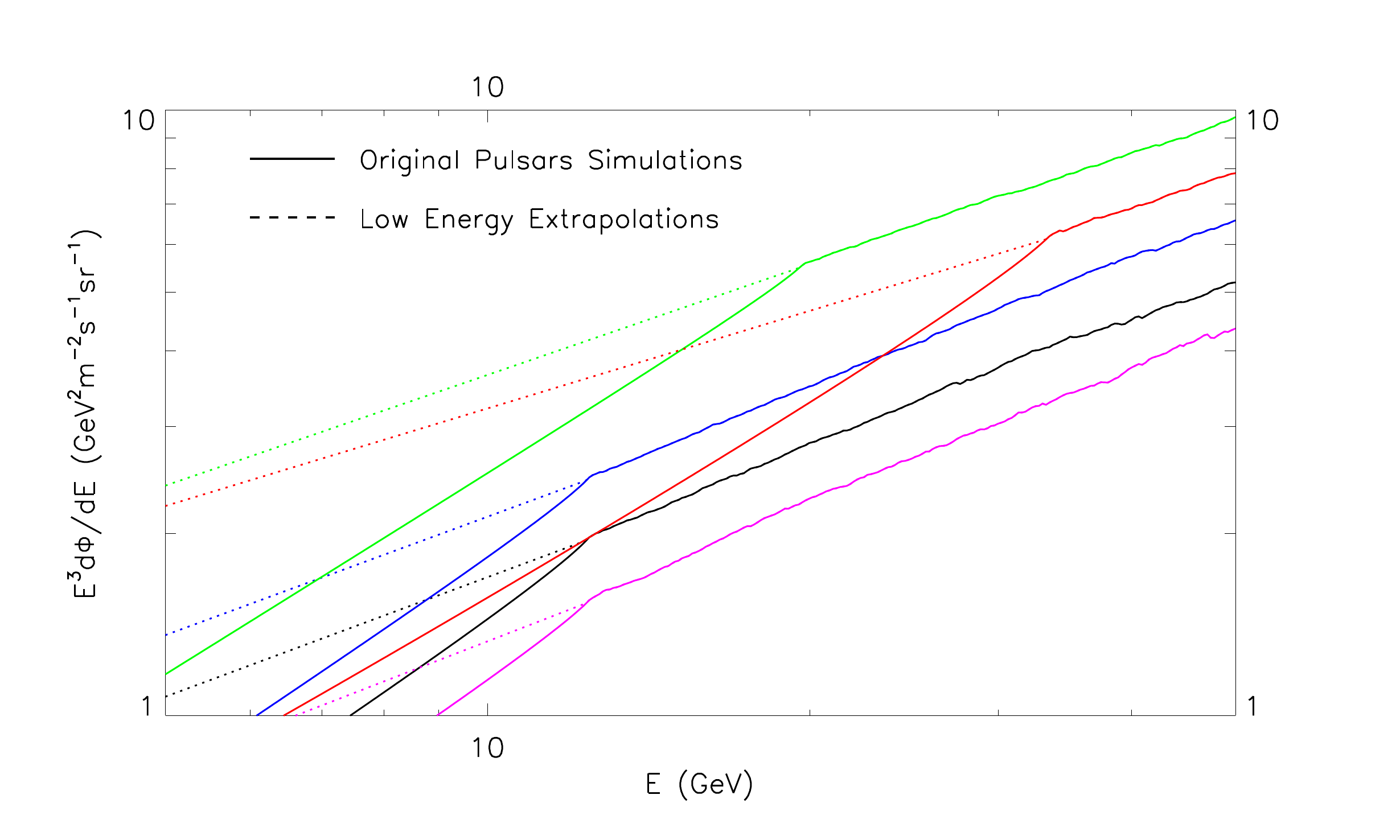}
\end{centering}
\vspace{-0.5cm}
\caption{Examples of low energy extrapolations on the $e^{\pm}$ flux from a sample of Milky Way
pulsars simulations. Normalizations are arbitrary and fluxes do not include the impact of solar modulation. 
For each original simulation (solid lines) we evaluate the power law before the break and then extrapolate 
to lower energies (dotted lines).}
\vspace{-0.4cm}
\label{fig:LowE_extrapol}
\end{figure}

For every Milky Way pulsars simulation we perform a low energy extrapolation.  
We show in Figure~\ref{fig:LowE_extrapol}, examples of that.
This extrapolation accounts for the fact that our simulations end at 4 kpc in distance from the Sun 
and at 10 Myr in age of pulsars. Through this extrapolation, more distant and older pulsars that would 
contribute at lower energies are included. The total number of Milky Way pulsars simulations that we
fit and compare to the observations is thus 14544 (instead of just 7272 simulations). However, we note
that simulations that can provide a good fit to the data both with and without their low-energy 
extrapolation, count only once in the list of simulations that are in agreement with the data. 

The exact ISM conditions beyond 4 kpc are not simulated for the pulsars' electron/positron flux components.  
As the ISM properties gradually change we expect that there is increased modeling uncertainly at the 
lower energies and both options should be considered viable. Moreover, at low energies we include in our
fits the impact of solar modulation with its uncertainties. This can further modify the fluxes' spectral properties
at these low energies. We note that while lower birth rates result in small features at the lower energies, 
this does not affect our ability to perform that 
extrapolation. That can be seen in Figure~\ref{fig:LowE_extrapol}, by comparing the blue vs black 
vs magenta lines that are for similar simulation assumptions, but with 2 vs 1 vs 0.6 pulsars/century 
birth rates respectively. The exact point where the low-energy extrapolation starts depends on the 
energy losses assumption. Lower energy losses (as the red line in Figure~\ref{fig:LowE_extrapol})
result in a higher energy from which the extrapolation starts. 

\begin{figure}[!]
\begin{centering}
\includegraphics[width=3.4in,angle=0]{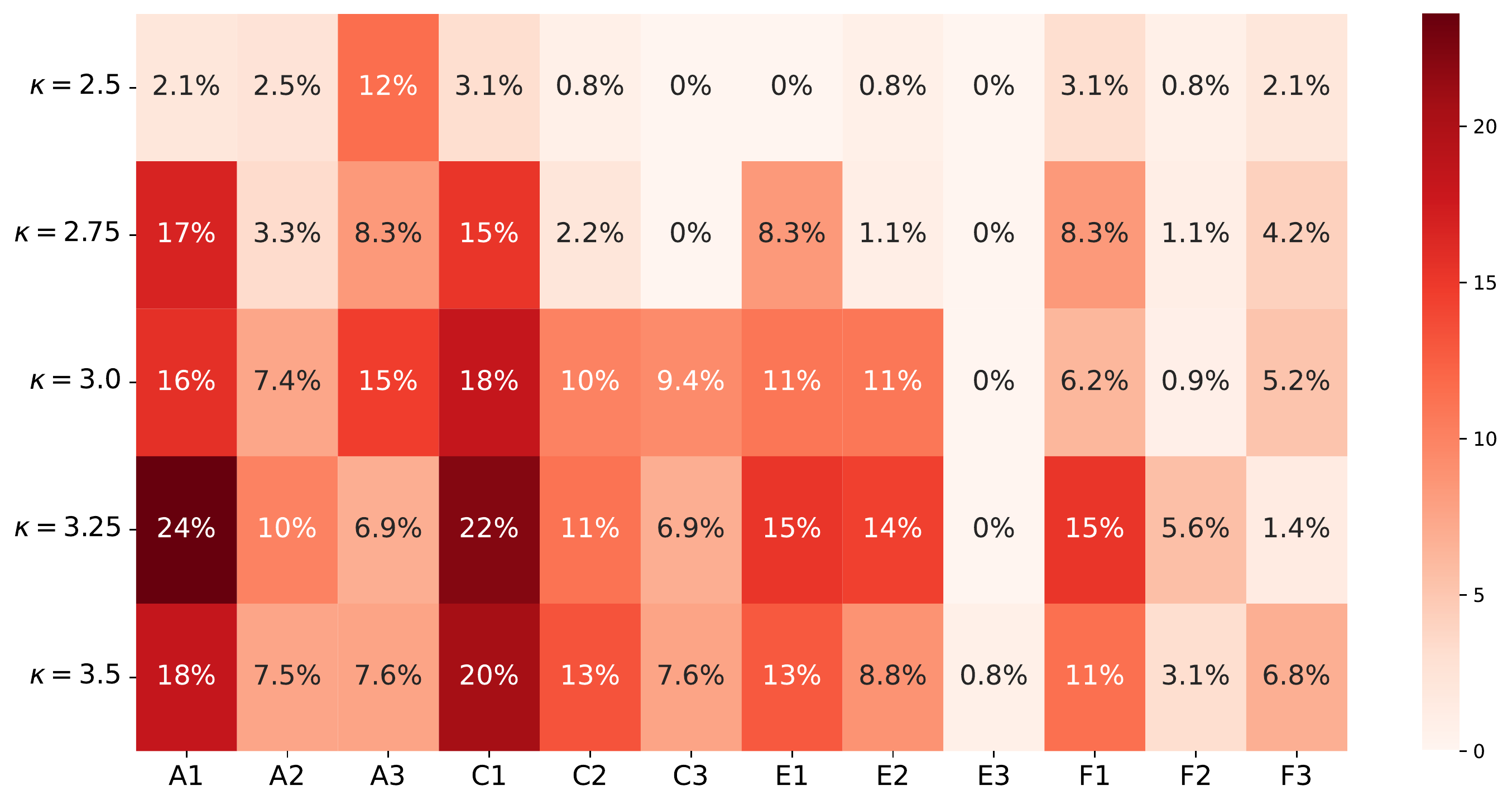}
\end{centering}
\vspace{-0.5cm}
\caption{In each cell we show the fraction of pulsar population simulations that are consistent within 2$\sigma$
to the \textit{AMS-02} positron fraction spectrum, the positron flux and the electron plus positron 
flux (see text for details) for the five choices of braking index $\kappa = 2.5, 2.75, 3.0, 3.25, 3.5$ 
and the twelve choices of ISM propagation conditions modeled by "A1" to  "F3" (see Table~\ref{tab:ISMBack}).}
\label{fig:Kappa_VS_ISM}
\end{figure}

\begin{figure}[!]
\begin{centering}
\includegraphics[width=3.4in,angle=0]{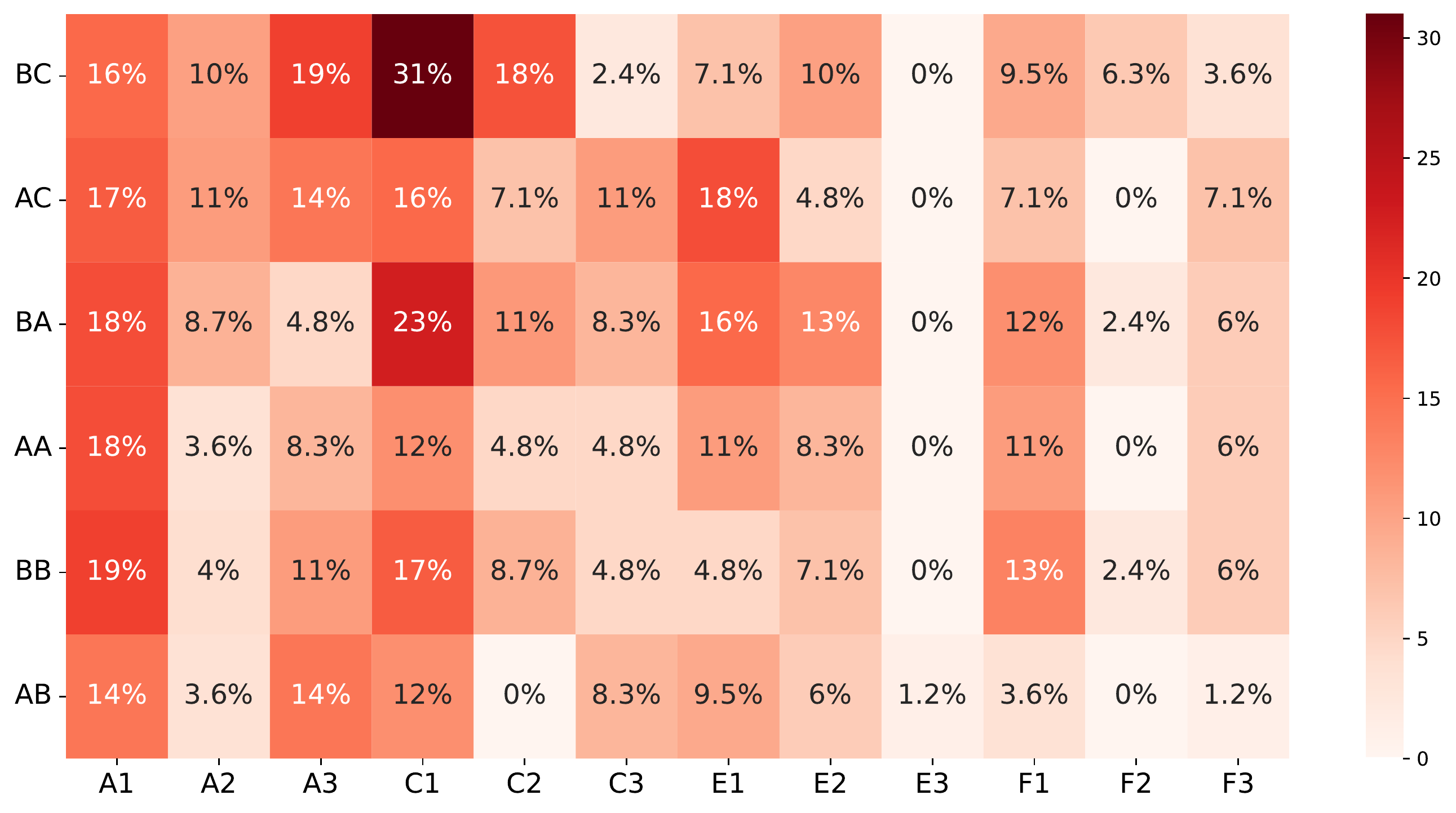}
\end{centering}
\vspace{-0.5cm}
\caption{Same as Figure~\ref{fig:Kappa_VS_ISM} for the
     combination of the six choices ("BC", "AC", "BA", "AA", "BB" and "AB")
     of $n$ and $g(\eta)$ and the twelve ISM models.}
\label{fig:Injec_VS_ISM}
\end{figure}

\begin{figure}[!]
\begin{centering}
\includegraphics[width=3.4in,angle=0]{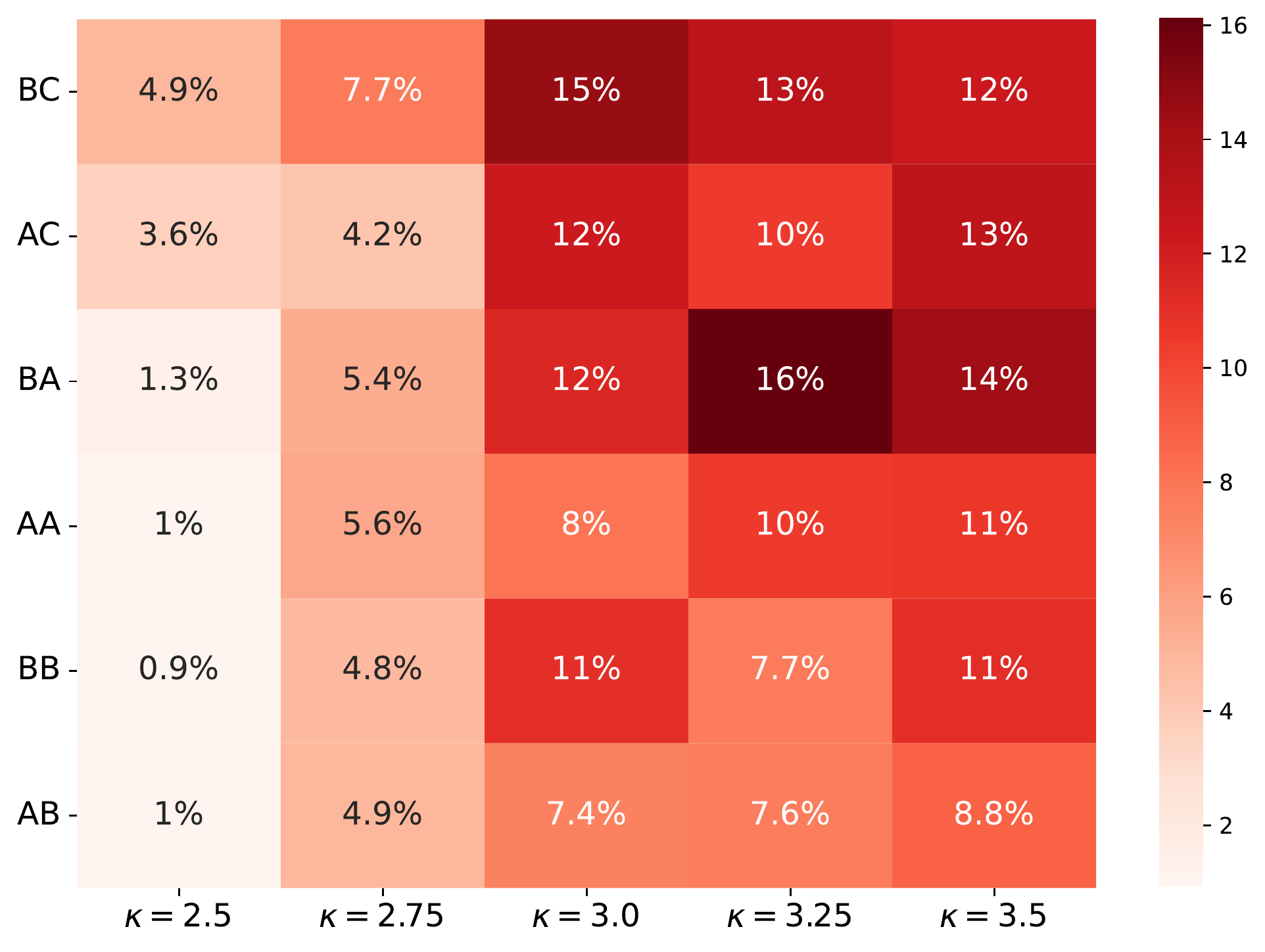}
\end{centering}
\vspace{-0.5cm}
\caption{Similar to Figures \ref{fig:Kappa_VS_ISM} and \ref{fig:Injec_VS_ISM},
     for the combination of the six choices of $n$
     and $g(\eta)$ and the five choices for braking index
     $\kappa = 2.5, 2.75, 3.0, 3.25, 3.5$.}
\label{fig:Injec_VS_Kappa}
\vspace{-0.4cm}
\end{figure}

Our findings on the pulsar properties and the ISM conditions can be summarized in Figures 
\ref{fig:Kappa_VS_ISM}, \ref{fig:Injec_VS_ISM} and  \ref{fig:Injec_VS_Kappa}.
We show in each cell the percentage of the pulsar population simulations that are consistent within 
$2\sigma$ to the \textit{AMS-02} positron fraction spectrum, the positron flux and electron plus positron flux.
For instance, the top left cell of Figure~\ref{fig:Kappa_VS_ISM} the value of $2.1$ refers to the percentage 
among the simulations with $\kappa=2.5$, ISM model A1 assumptions, that are consistent with 
the \textit{AMS-02} data within $2\sigma$.

In Figure~\ref{fig:Kappa_VS_ISM}, we show our results for the combination of the five choices 
of the braking index and the twelve choices of ISM propagation conditions.
The choices of braking index $\kappa=2.5$ and $\kappa=2.75$ are clearly disfavored. 
Only a few realizations with these choices survive within $2\sigma$. For $\kappa=3.0$ 
and even more so for $\kappa=3.25$ there is a significant increase on the fraction of 
simulations consistent to the data. Simulations with $\kappa = 3.5$ are favored as much
as simulations with $\kappa = 3.0$. The preference for $\kappa \ge 3.0$ can also be seen
in Figure~\ref{fig:Injec_VS_Kappa}. In Ref.~\cite{Cholis:2018izy}, a first indication for a 
$\kappa \ge 3.0$ tendency was found. We now confirm this tendency with a much larger 
set of simulations, that account for even greater range of the relevant parameter space 
being modeled. For the $\sim 10$ young pulsars that we have reliable measurements of 
their breaking index, typical values are $\kappa < 3$ \cite{Hamil:2015hqa, Archibald:2016hxz}.  
A higher value than 3.0 for the braking index, represents a slower spin-down 
(see Eq.~{\ref{eq:SpinDown}) for middle aged and multi-Myr old pulsars. While the sample
of young pulsars is still small, one solution between the results from electromagnetic 
observations and the results of this analysis is that pulsars may increase their braking as 
they age. As we rely on observations from radio waves and evolve the pulsars back in time to
calculate their total power, a constant braking index of $\kappa < 3$, produces pulsars that initially 
were very powerful sources emitting very large amounts of cosmic-ray electrons and positrons. 
This increases the emission from the older pulsars compared to the younger ones.  
  
As can be seen from both Figures~\ref{fig:Kappa_VS_ISM} and~\ref{fig:Injec_VS_ISM} there
is a small preference for the propagation models "A1, "C1", "E1" and "F1" which are for the
more commonly used assumptions on the local cosmic-ray electrons and positron losses . 
Interestingly, practically 0$\%$ of our astrophysical realizations that have propagation 
model "E3" are consistent to the data. The "E3" simulations model low-energy losses and 
fast diffusion of high-energy cosmic rays. Also, models for a thin diffusion disk ("F1", "F2" and 
"F3") are disfavored. Thin disk ISM models make the cosmic-ray positrons even at low energies 
quite local and suppress the overall pulsars' contribution, making the combined spectrum  from 
pulsars harder, something that is in tension to the observations. 

In Figure~\ref{fig:Injec_VS_ISM}, we present the results for the combination of the six choices
for the injection index $n$ and the energy conversion to cosmic-rays $g(\eta)$ ("BC", "AC", 
"BA", "AA", "BB" and "AB") and the twelve ISM models. We clarify that there are three more 
choices ("CA", "CB" and "CC") for $n \in [1.3, 1.5]$ that are not shown here. These would be 
redundant rows in the table as none of the realizations with these choices are within our 
$2\sigma$ limit. The preferences on the ISM properties are there as we noted before. Regarding the 
index $n$ described by the first letter ("A" or "B") along the rows, there a slight preference for 
the choice "B" which is for a narrow range of values for $n \in [1.6, 1.7]$ over the choice 
"A", which is for a wider range of $n \in [1.4, 1.9]$ (see Section~\ref{subsec:PulsarSources} 
for further details). The broader range for $n$ (under choice "A") results in a more diverse pulsar 
population with respect to their injected cosmic-ray electrons and positrons. Therefore pulsars 
under assumption "A" have quite diverse spectra. The resulting combined pulsars' spectrum in 
turn has some very pronounced features associated with pulsars that have $n$ closer to 1.4. Such 
a choice while not fully excluded is less preferable. Choice "C" which is fully excluded, assumed 
that all pulsars have values of $n \in [1.3, 1.5]$. This resulted in spectra with too many strong spectral 
features compared to what is observed in the \textit{AMS-02} data. Our results suggest that 
pulsars in the Milky Way most likely have a small range of values for their spectral index $n$ with
values of $n \simeq 1.6$.

In Figure~\ref{fig:Injec_VS_ISM}, when comparing our results with respect to the choices of $g(\eta)$ 
(depicted by the second letter along the rows), we see a gradient going from our choice "C", to choice 
"A" and then to "B". Those are ranked from smaller to larger variance on the fraction of power $\eta$ that 
goes to cosmic rays (see Section~\ref{subsec:PulsarSources}). Our results suggest that pulsars 
simulations with a greater homogeneity in the total spin-down power converted to comic-rays are preferred. If 
pulsars had a very large range of $\eta$, then even among the older pulsars contributing at the lower 
energies, we would have a few of them standing out in their produced fluxes. Those pulsars would 
again give more spectral features than what is observed. 

In Figure~\ref{fig:Injec_VS_Kappa}, we present our results for the combination of the six choices 
for $n$ and $g(\eta)$, and the five choices for braking index. This figure, is given as a projection 
along that part of the simulations parameter space and very clearly shows the preferences for 
both $\kappa \ge 3.0$ and the choice "B", to the "A" for $n$.  

\subsubsection{Observations at TeV energies from \textit{DAMPE} and \textit{CALET}}

The positron flux is measured by \textit{AMS-02} up to energies of 1 TeV. Two more experiments, 
\textit{DAMPE} \cite{TheDAMPE:2017dtc} and \textit{CALET} \cite{2015JPhCS.632a2023A} have 
published their measurements of the total $e^{+} + e^{-}$ CR flux up to 5 TeV \cite{DAMPE:2017fbg, 
Adriani:2018ktz}. At these energies the pulsars' contribution to the total measured fluxes can be 
very significant. The combination of volume and age necessary for pulsars to be able to contribute, 
is reduced at these higher energies. We can only probe the properties of the youngest and 
most energetic pulsars that are also close-by members of the population. Those are small in number.
The result is an $e^{+} + e^{-}$ flux rich in spectral features. Such spectra can be seen in 
Figures~\ref{fig:DampeLeptons} and \ref{fig:CaletLeptons} where we show the predicted $e^{+} + e^{-}$ 
fluxes for some of our pulsars models. Above 1 TeV, the spectra can either have a cut-off, a 
change in their slope, or in some cases one or more prominent bumps from individual pulsars.

\begin{figure}
\begin{centering}
\hspace{-0.4cm}
\includegraphics[width=3.80in,angle=0]{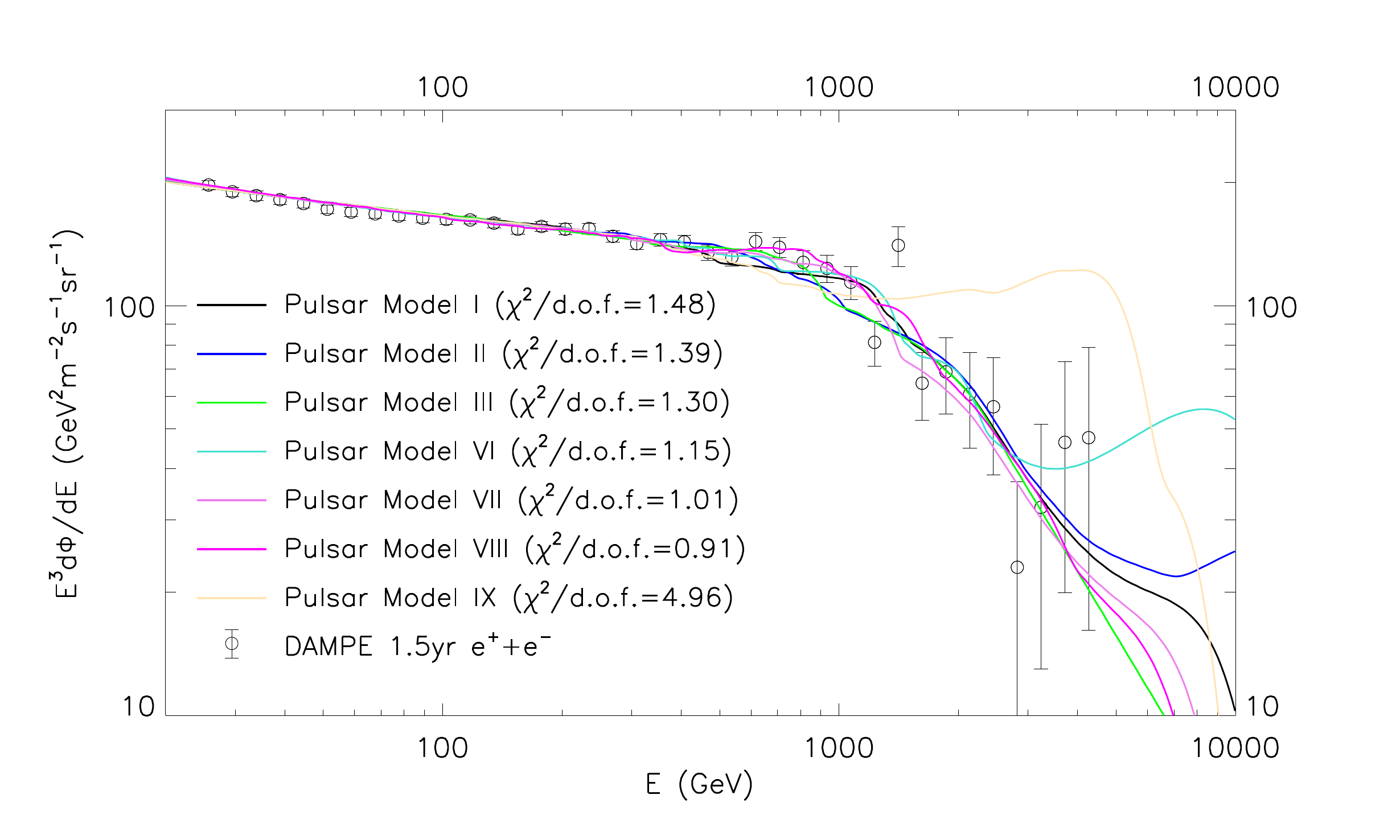}
\end{centering}
\vspace{-0.9cm}
\caption{The predicted $e^+ + e^-$ spectra for seven different pulsar models along with their $\chi^2/$d.o.f. 
One of the models (model IX in beige color) is excluded by the \textit{DAMPE} data. This model 
greatly overshoots the data points above 1 TeV. Some fluxes (such as model VI and VIII) contain notable 
features at TeV energies from individual powerful pulsars.} 
\vspace{-0.4cm}
\label{fig:DampeLeptons}
\end{figure}

Fitting our simulations to the \textit{DAMPE} $e^{+} + e^{-}$ spectrum, we can further constrain the local 
Milky Way pulsars properties. Of the 567 (1095) realizations that were within $2\sigma  \,(3\sigma)$ to all 
three \textit{AMS-02} datasets, 268 (771) are also consistent within $2\sigma \, (3\sigma)$ to the \textit{DAMPE} 
$e^{+} + e^{-}$ data. 
\textit{DAMPE} allows roughly three quarters of our models at the $3\sigma$ level and roughly  half of our models at the $2\sigma$ level.
We note that \textit{DAMPE} excludes models in a uniform manner from the heat maps  of Figures~{\ref{fig:Kappa_VS_ISM}, 
\ref{fig:Injec_VS_ISM} and \ref{fig:Injec_VS_Kappa}. Thus our conclusions on the averaged properties of 
pulsars and the local ISM do not change. One difference is that with \textit{DAMPE} we find a small preference 
in retaining simulations with lower birth rates suggestive of the fact that smaller rates more easily produce 
spectral features as the ones seen in the \textit{DAMPE} data.
However, realizations that contain extremely powerful young and near-by pulsars can cause the $e^+ + e^-$ 
flux to overshoot the \textit{DAMPE} data at the highest energies and are excluded. In Figure~\ref{fig:DampeLeptons}, 
we show the predicted $e^{+} + e^{-}$ flux from six pulsar models that are consistent with the \textit{DAMPE}.
We also show the flux from one model (Pulsar Model IX) that is not consistent with the \textit{DAMPE} data 
exactly due to the presence of very powerful pulsars at energies that \textit{AMS-02} can not measure. 

\begin{figure}
\begin{centering}
\hspace{-0.4cm}
\includegraphics[width=3.80in,angle=0]{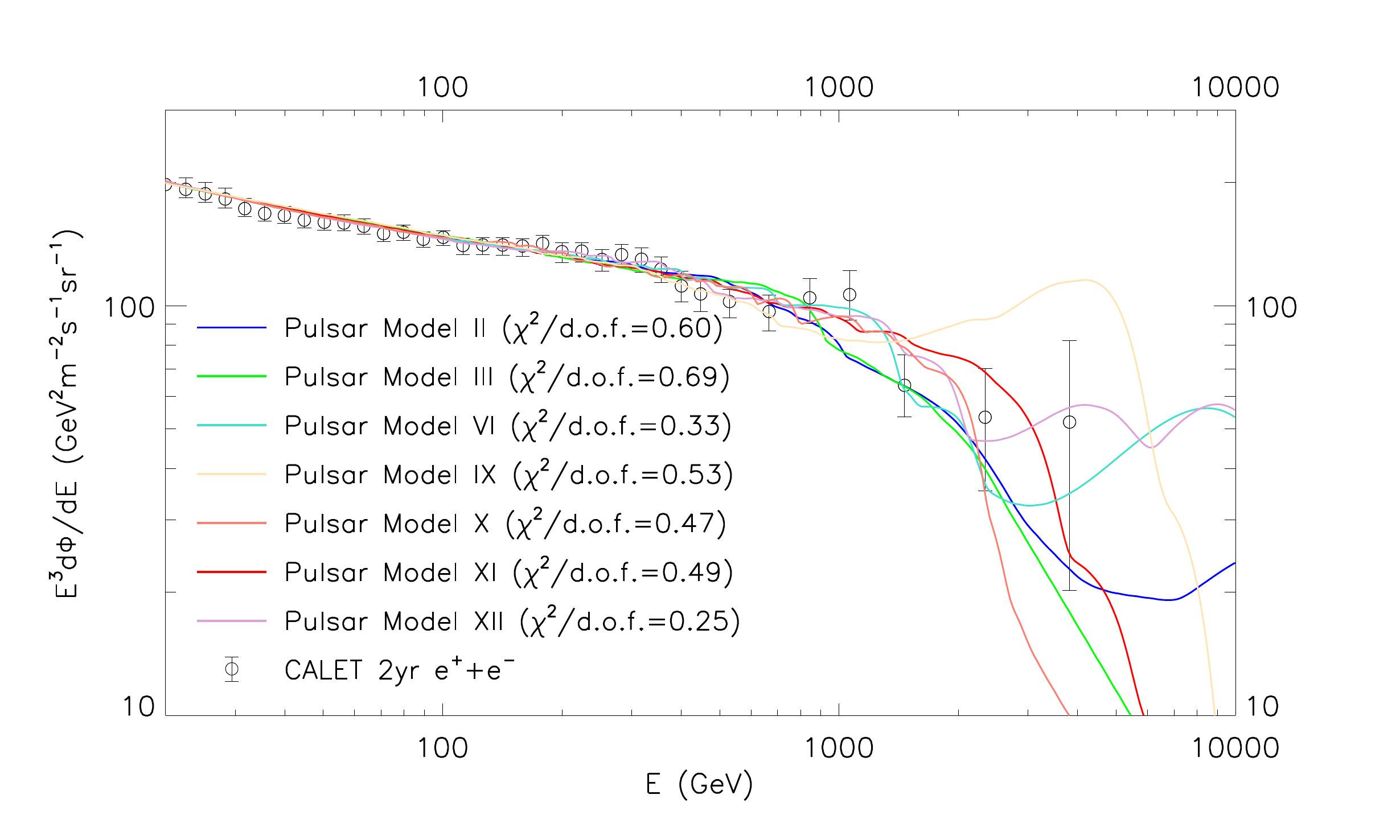}
\end{centering}
\vspace{-0.9cm}
\caption{Seven different predicted $e^{+} + e^{-}$ fluxes from realizations that can fit the \textit{CALET} data 
and their  $\chi^2/\textrm{d.o.f.}$. Pulsar Model IX, which was excluded by \textit{DAMPE} in Figure 
\ref{fig:DampeLeptons} can fit the \textit{CALET} data very well. Prominent features are also visible 
at the highest energies from powerful pulsars (models VI, IX, XI and XII).} 
\vspace{-0.4cm}
\label{fig:CaletLeptons}
\end{figure}

\textit{CALET}'s observations can not further constrain our astrophysical realizations. Of the 567 (1095) 
realizations that were consistent with all \textit{AMS-02} data at $2\sigma$ ($3\sigma$), only 
one is excluded by the \textit{CALET}  $e^{+} + e^{-}$ flux.  \textit{CALET} has larger error bars 
compared to \textit{DAMPE}. Almost all of our 1095  realizations that are within 
$3\sigma$ to the \textit{AMS-02} measurements, end up with $\chi^2/\textrm{d.o.f.}<1$ fit to the 
\textit{CALET} data.
In Figure~\ref{fig:CaletLeptons}, we show seven different $e^{+}+ e^{-}$ fluxes from realizations 
that fit the \textit{CALET} observation. 
Even our model IX, that was excluded by \textit{DAMPE} with $\chi^2/\textrm{d.o.f.}=4.96$, 
can fit the \textit{CALET} data with $\chi^2/\textrm{d.o.f.}=0.53$. 

Finally, we use the fits to the \textit{DAMPE} and \textit{CALET} in combination with the \textit{AMS-02} fits, 
to test the overall conversion efficiency $\eta$ of pulsars' spin-down power to power in cosmic-ray electrons 
and positrons. In Figure~\ref{fig:Eta_dist}, we show for the allowed pulsars simulations their fitted mean 
efficiency $\bar{\eta}$. We find that $\bar{\eta}$ is typically between  0.05 and 0.2. Due to equipartition 
of energy in the pulsar's environments, we do not allow for $\bar{\eta} > 0.5$ in our fits. This is also shown 
in Figure~\ref{fig:Eta_dist}. There is only a small number of pulsar simulations with $\bar{\eta} < 0.01$. 
Recently, an independent analysis of \cite{Orusa:2021tts} suggested a similar fraction of spin-down power
to cosmic-ray electrons and positrons.
We present the simulations that are within the $2\sigma$ criterion, but note that even if we used the larger 
number of simulations that are within $3\sigma$, the results would not differ. 
\begin{figure}
\begin{centering}
\hspace{-0.6cm}
\includegraphics[width=3.72in,angle=0]{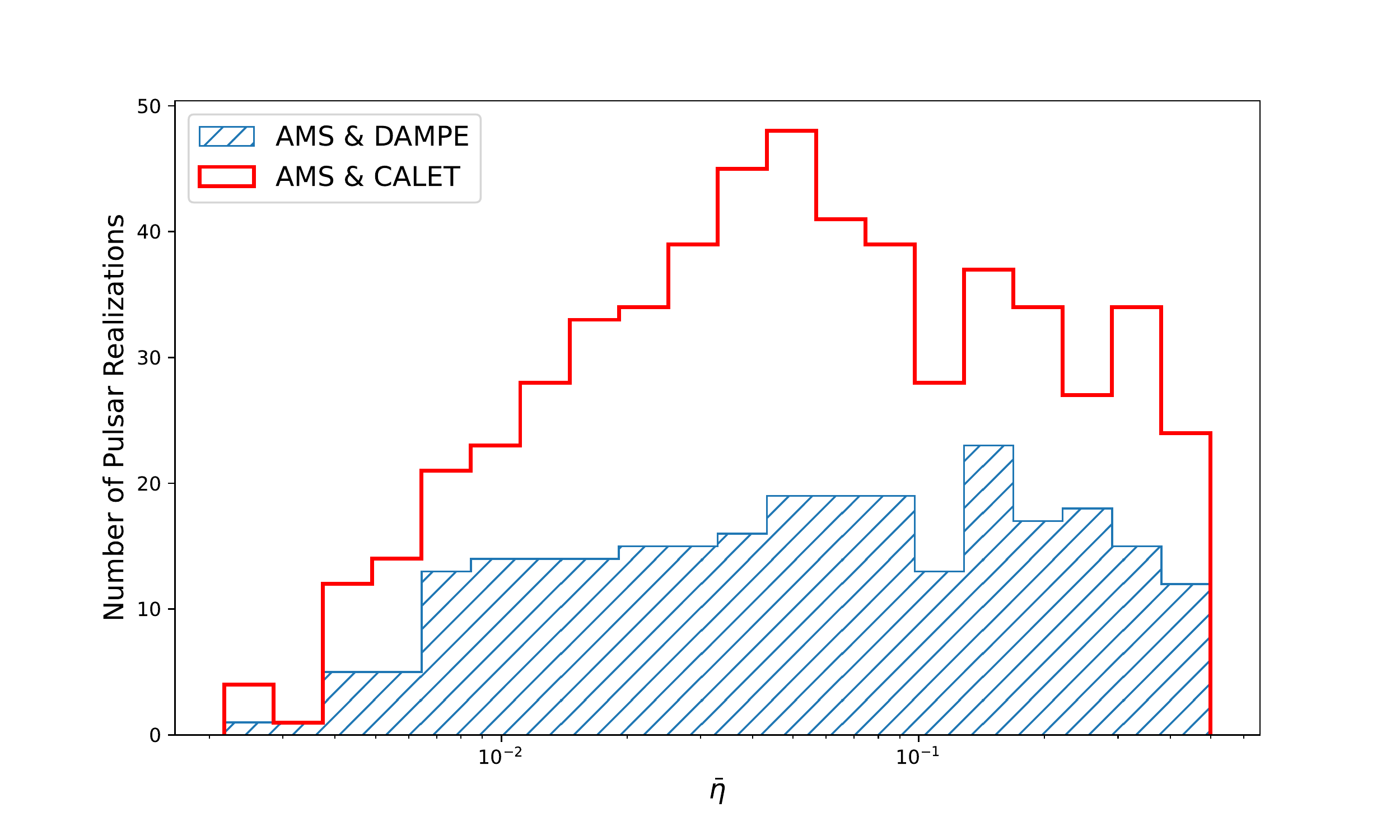}
\end{centering}
\vspace{-0.9cm}
\caption{The distribution of the fitted values for the averaged conversion efficiency $\bar{\eta}$ of pulsars' 
spin down power to cosmic rays. The y-axis shows the number of allowed Milky Way pulsars simulations within 
$2\sigma$. Typical values for $\bar{\eta}$ are $\simeq 0.1$.}
\vspace{-0.4cm}
\label{fig:Eta_dist}
\end{figure}

\subsection{Results for $E \ge 5\ \textrm{GeV}$ fits}
\label{subsec:Results5GeV}

In this section we include the \textit{AMS-02} positron fraction and positron flux measurements down to 
energies of 5 GeV and discuss the relevant implications of the increased energy range. For the total 
$e^{+} + e^{-}$ flux we still fit the data above 15 GeV. The presence of the feature at $\simeq$ 12 GeV has an 
effect in the results of this section. Instead, in the discussion of Section~\ref{subsec:Results15GeV}, 
we fitted energies $E > 15$ GeV, avoiding its impact on the Milky Way pulsar properties. 

As we have discussed also in Section~\ref{subsec:Results15GeV}, for each pulsars simulation, 
we consider a low energy extrapolation as well. We find that this low-energy extrapolation being 
included is more important than in Section~\ref{subsec:Results15GeV}, 
where in fact for some ISM energy-loss assumptions the extrapolation starts at energies lower than 
15 GeV. Even with the increased level of allowed low-energy flux uncertainty, many more simulations 
can be excluded by the data than in Section~\ref{subsec:Results15GeV}. That is to be expected 
as the \textit{AMS-02} positron and electron fluxes at these energies have significantly smaller statistical errors. 
However, as we will show the presence of the feature around 12 GeV also has an impact. Including 
the positron fraction and positron flux measurements from 5 to 15 GeV, only a handful 
of Milky Way pulsars simulations are within $2\sigma$ agreement to the \textit{AMS-02} observations. 
For the remainder we focus on simulations that are within $3\sigma$ and $5\sigma$ from an 
expectation of $\chi^2$ of 1 per degree of freedom, i.e. simulations that are not excluded 
within $3\sigma$ and $5\sigma$. To demonstrate how more challenging it is to fit the \textit{AMS-02} 
data at $E \ge 5$ GeV compared to $E > 15$ GeV we include Figure~\ref{fig:ComparisonHeatmaps}.
On the left part of that figure, we show the fraction of allowed simulations within $3\sigma$ from the
fits to $E>15$ GeV and on the right the equivalent fraction of allowed simulations within $3\sigma$ 
from the fits to $E \ge 5$ GeV. There is a dramatic decrease in the number of allowed simulations.
\begin{figure*}
\begin{centering}
\hspace{-0.1cm}
\includegraphics[width=3.50in,angle=0]{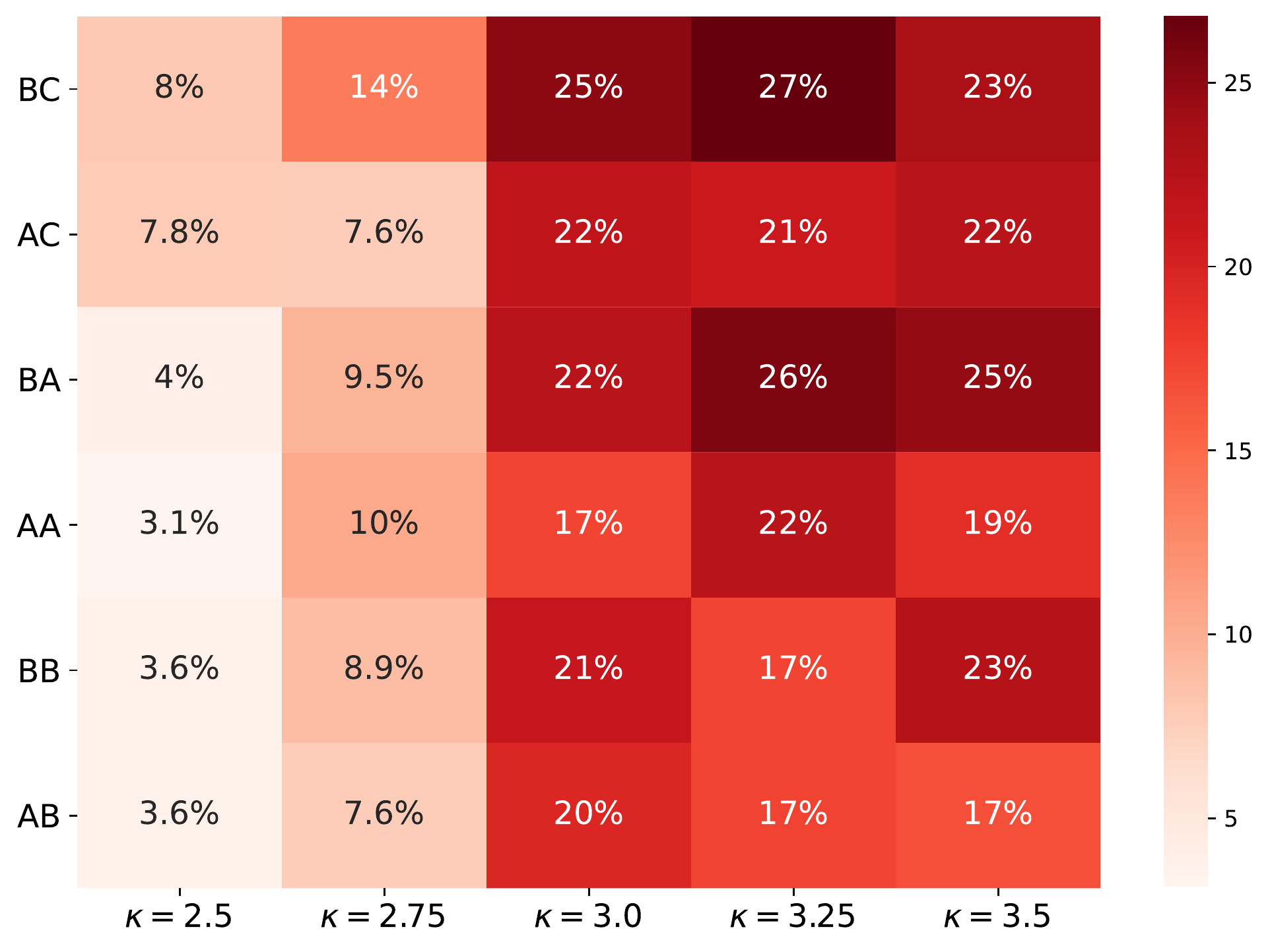}
\includegraphics[width=3.50in,angle=0]{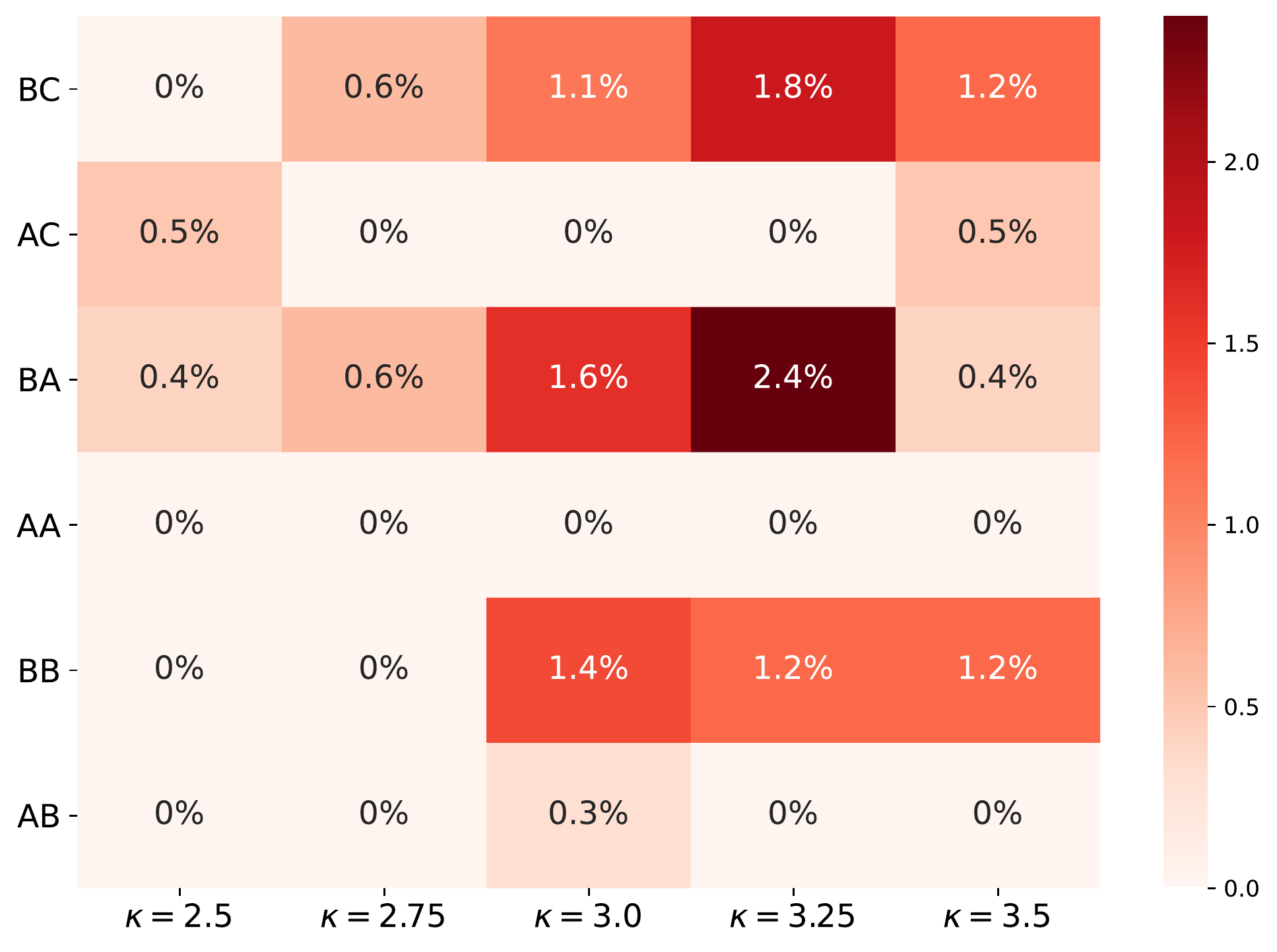}
\end{centering}
\vspace{-0.5cm}
\caption{Similarly to Figure~\ref{fig:Injec_VS_Kappa}, for given combination of pulsars' properties 
(each cell), we show the fraction of pulsar population simulations that are consistent within 3$\sigma$.
On the left we have fitted the simulations to the \textit{AMS-02} observations from $E > 15$ GeV, 
and on the right to the \textit{AMS-02} observations from $E \ge 5$ GeV (see text for further details).}
\vspace{-0.0cm}
\label{fig:ComparisonHeatmaps}
\end{figure*}

We test separately our simulations to each of the three 
\textit{AMS-02}  measurements and present results that are consistent with all three 
(for more details see Section~\ref{sec:Fitting}). We find that of the 7272 astrophysical 
realizations, 2831 can fit the \textit{AMS-02} positron-flux spectrum within $5\sigma$. Of these 2831 
realizations, only 37 (261) can also fit within $3\sigma \; (5\sigma)$ the \textit{AMS-02} positron 
fraction spectrum and the $e^{+} + e^{-}$ spectrum. The positron fraction has by far the greatest impact in 
excluding simulations \footnote{Just fitting the the positron 
fraction, we find only 50 (325) pulsars simulations to it within $3\sigma \, (5\sigma)$}. 

We summarize in tables \ref{tab:Kappa_VS_ISM}, \ref{tab:Injec_VS_ISM} and  \ref{tab:Injec_VS_Kappa} the 
properties of the pulsars and the ISM needed to explain the \textit{AMS-02} observations. For every cell,
we show the percentage of the pulsars simulations that are consistent within $3\sigma$ and $5\sigma$ 
(in parentheses) to the \textit{AMS-02} data. For instance, in Table \ref{tab:Kappa_VS_ISM} of the 
simulations with the combination of $\kappa=3.0$ and ISM assumptions "A2", only 2$\%$ (15$\%$) 
are consistent within $3\sigma$ ($5\sigma$) to the \textit{AMS-02} data. 
\begin{table*}[t]
    \begin{tabular}{c|cccccccccccc}
         \hline
             & A1 & A2 & A3 & C1 & C2 & C3 & E1 & E2 & E3 & F1 & F2 & F3 \\
             & ($\%$) & ($\%$) & ($\%$) & ($\%$) & ($\%$) & ($\%$) & ($\%$) & ($\%$) & ($\%$) & ($\%$) & ($\%$) & ($\%$) \\
             \hline \hline
            $\kappa = 2.5$ & 0 (0) & 0.8 (4) & 0 (0) & 0 (0) & 0 (0.8) & 0 (0) & 0 (0) & 0.8 (2) & 0 (0) & 0 (0) & 0 (2) & 0 (0) \\
            $\kappa = 2.75$ & 0 (0) & 0 (7) & 0 (0) & 0 (0) & 1 (3) & 0 (0) & 0 (1) & 0 (2) & 0 (0) & 0 (0) & 1 (6) & 0 (0) \\
            $\kappa = 3.0$ & 0 (1) & 2 (15) & 0 (0) & 0 (0) & 2 (11) & 0 (0) & 0 (0) & 3 (11) & 0 (0) & 0 (0) & 0.4 (4) & 0 (0) \\
            $\kappa = 3.25$ & 0 (1) & 1 (16) & 0 (0) & 0 (3) & 6 (18) & 0 (0) & 0 (1) & 1 (20) & 0 (0) & 0 (0) & 2 (10) & 0 (0) \\
            $\kappa = 3.5$ & 0 (0) & 1 (11) & 0 (0) & 0 (0) & 3 (15) & 0 (0) & 0 (0) & 0.6 (11) & 0 (0) & 0 (0) & 0 (8) & 0 (0) \\
            \hline \hline
        \end{tabular}
\caption{We show our results for the combination of the five choices of braking index $\kappa = 2.5, 2.75, 3.0, 3.25, 3.5$ 
and the twelve choices of ISM propagation conditions modeled by "A1" to  "F3" (see Table~\ref{tab:ISMBack}). We give 
the fraction of pulsar population simulations that are consistent within 3$\sigma$ and 5$\sigma$ limits (in parentheses)
to the \textit{AMS-02} positron fraction spectrum, the positron flux and the electron+positron flux (see text for details).
For the combination of $\kappa$ = 3 and "C2" we produced 288 simulations to probe the remaining astrophysical parameters, 
of which 5 (31) i.e. $\simeq$2$\%$ (11$\%$) are allowed within 3$\sigma$ (5$\sigma$).}
\vspace{-0.2cm}
\label{tab:Kappa_VS_ISM}
\end{table*}

\begin{table*}[t]
    \begin{tabular}{c|c|cccccccccccc}
         \hline
             & & A1 & A2 & A3 & C1 & C2 & C3 & E1 & E2 & E3 & F1 & F2 & F3  \\
             & & ($\%$) & ($\%$) & ($\%$) & ($\%$) & ($\%$) & ($\%$) & ($\%$) & ($\%$) & ($\%$) & ($\%$) & ($\%$) & ($\%$) \\
            \hline \hline
            BC & $1.6 \leq n \leq 1.7$, $\zeta = 1.29$   & 0 (0) & 3 (18) & 0 (0) & 0 (0) & 3 (19) & 0 (0) & 0 (0) & 0.8 (9) & 0 (0) & 0 (0) & 2 (14) & 0 (0) \\
            AC & $1.4 \leq n \leq 1.9$, $\zeta = 1.29$   & 0 (2) & 0 (13) & 0 (0) & 0 (0) & 1 (8) & 0 (0) & 0 (0) & 1 (4) & 0 (0) & 0 (0) & 0 (4) & 0 (0) \\
            BA & $1.6 \leq n \leq 1.7$, $\zeta = 1.47$   & 0 (0) & 2 (19) & 0 (0) & 0 (0) & 5 (15) & 0 (0) & 0 (0) & 2 (20) & 0 (0) & 0 (0) & 2 (6) & 0 (0) \\
            AA & $1.4 \leq n \leq 1.9$, $\zeta = 1.47$   & 0 (1) & 0 (10) & 0 (0) & 0 (0) & 0 (8) & 0 (0) & 0 (0) & 0 (10) & 0 (0) & 0 (0) & 0 (2) & 0 (0) \\
            BB & $1.6 \leq n \leq 1.7$, $\zeta = 2.85$   & 0 (0) & 2 (7) & 0 (0) & 0 (0) & 3 (13) & 0 (0) & 0 (2) & 2 (10) & 0 (0) & 0 (0) & 0 (4) & 0 (0) \\
            AB & $1.4 \leq n \leq 1.9$, $\zeta = 2.85$   & 0 (0) & 0 (2) & 0 (0) & 0 (2) & 0 (2) & 0 (0) & 0 (0) & 1 (7) & 0 (0) & 0 (0) & 0 (1) & 0 (0) \\
            \hline \hline 
        \end{tabular}
\caption{As in Table~\ref{tab:Kappa_VS_ISM}, we present, for the
     combination of the six choices ("BC", "AC", "BA", "AA", "BB" and "AB")
     of $n$ and $g(\eta)$ and the twelve ISM models, the $\%$
     fraction of pulsar simulations that are consistent within the
     3$\sigma$ and 5$\sigma$ limits (in parentheses) to the \textit{AMS-02} cosmic-ray data.} 
\vspace{-0.1cm}
\label{tab:Injec_VS_ISM}
\end{table*}

\begin{table*}[t]
    \begin{tabular}{c|c|ccccc}
         \hline
             & & $\kappa = 2.5$ & $\kappa = 2.75$ & $\kappa = 3.0$ & $\kappa = 3.25$ & $\kappa = 3.5$ \\
             & & ($\%$) & ($\%$) & ($\%$) & ($\%$) & ($\%$) \\
            \hline \hline
            BC & $1.6 \leq n \leq 1.7$, $\zeta = 1.29$ & 0 (1) & 0.6 (3) & 1 (8) & 2 (10) & 1 (8) \\
            AC & $1.4 \leq n \leq 1.9$, $\zeta = 1.29$ & 0.5 (1) & 0 (1) & 0 (4) & 0 (2) & 0.5 (3) \\
            BA & $1.6 \leq n \leq 1.7$, $\zeta = 1.47$ & 0.4 (3) & 0.6 (4) & 2 (7) & 2 (12) & 0.4 (7) \\
            AA & $1.4 \leq n \leq 1.9$, $\zeta = 1.47$ & 0 (0) & 0 (0.7) & 0 (4) & 0 (8) & 0 (0.9) \\
            BB & $1.6 \leq n \leq 1.7$, $\zeta = 2.85$ & 0 (0) & 0 (2) & 1 (4) & 1 (4) & 1 (8) \\
            AB & $1.4 \leq n \leq 1.9$, $\zeta = 2.85$ & 0 (0) & 0 (0) & 0.3 (2) & 0 (2) & 0 (2) \\
            \hline \hline
        \end{tabular}
\caption{Similar to the slices in parameter space given in
     Tables~\ref{tab:Kappa_VS_ISM} and \ref{tab:Injec_VS_ISM},
     we show for the combination of the six choices of $n$
     and $g(\eta)$ and the five choices for braking index
     $\kappa = 2.5, 2.75, 3.0, 3.25, 3.5$, the $\%$ fraction of pulsar population
     simulations that are consistent within 3$\sigma$ and 5$\sigma$ to the \textit{AMS-02} cosmic-ray data.}
\vspace{-0.1cm}
\label{tab:Injec_VS_Kappa}
\end{table*}

In Table \ref{tab:Kappa_VS_ISM}, we show our results for the combination of the five 
choices of the braking index and the twelve choices of ISM propagation conditions.
There is a clear preference for the ISM models "A2", "C2", "E2" and "F2" that predict
the highest energy losses ($b=8.02\times 10^{-6}$GeV$^{-1}$kyrs$^{-1}$ in Table 
\ref{tab:ISMBack}). That strong preference is mostly the result of a spectral feature 
at $\simeq$ 12 GeV. The "A2", "C2", "E2" and "F2" without their low-energy extrapolation have a
break close to that energy. We believe this feature likely suggests an additional source
of positrons around 10-15 GeV, which may be a population of more distant pulsars 
closer to the inner spiral arm, around 5 kpc away from us. A second reason why low ISM
energy losses models perform worse once including the 5-15 GeV data to our fits, is that 
these simulations predict fluxes that overshoot the observed spectra at high energies. 

Our finding from Section~\ref{subsec:Results15GeV} for a preference of a braking index 
$\kappa \ge 3.0$ remains to be the case even with the lower energy observations. In Table 
\ref{tab:Kappa_VS_ISM}, the choices $\kappa=2.5$ and $\kappa=2.75$ are almost entirely 
excluded  (see also Table \ref{tab:Injec_VS_Kappa}). Instead, for $\kappa=3.0$ there is a 
significant increase in the percentage of simulations in agreement to the data, followed by 
an even higher percentage for $\kappa=3.25$. For $\kappa=3.5$ we find similar results as 
for $\kappa=3.0$. 

In Table \ref{tab:Injec_VS_ISM}, we present as in Figure~\ref{fig:Injec_VS_ISM} our the results for the combination 
of injection index $n$ and $g(\eta)$ and the twelve ISM models. Our findings are similar to those 
presented in Section~\ref{subsec:Results15GeV}, with the difference that now there are practically no
simulations allowed with $n \in [1.4, 1.9]$ at $3 \sigma$ (first letter "A" along the raws). Again 
the choices of "CA", "CB" and "CC" are not presented as only one simulation was found to be within 
$5\sigma$ to the data. 
Thus in our analysis, we can probe effectively the distribution properties of the injection index $n$. 
Yet, the stronger degeneracies on $g(\eta)$ are more difficult to further reduce with the lower energy data. 
At these low energies we observe the combined fluxes from thousands of pulsars and also the uncertainties
associated with the cosmic ray secondaries become prominent. We note that the similarity on the 
derived properties of $n$ and $g(\eta)$ in the results of this section and Section~\ref{subsec:Results15GeV}
are to be expected. Varying the assumptions on $n$ and $g(\eta)$ affects the shape and magnitude
of spectral features appearing at high energies and not between 5 and 15 GeV.

In Table \ref{tab:Injec_VS_Kappa}, we present for completeness our results for the combination of 
the six choices for $n$ and $g(\eta)$, and the five choices for braking index. The domination of the 
choice "B", followed by choice "A", for $n$ that was described in the previous paragraphs is shown, 
and also the preference for braking index values $\kappa \ge 3$. 

\section{Discussion and Conclusions}
\label{sec:Conclusions}

In this paper we use recent cosmic-ray electron and positron observations from the \textit{AMS-02}, 
\textit{DAMPE} and \textit{CALET} collaborations as a new handle to constrain 
the properties of Milky Way pulsars. Unlike electromagnetic spectrum observations that study individual
objects, in our work we do not constrain the properties of any single pulsar. Instead, we constrain 
the properties of the general population of pulsars that can contribute to the observed cosmic-ray fluxes 
from 5 GeV up to 5 TeV. As the observed electron/positron cosmic-ray energy increases the volume of 
possible sources decreases. This makes our analysis restricted to a smaller fraction of the local ISM 
volume at the highest energies. 

We created simulations of the Milky Way pulsars that lie within 4 kpc from the location of the Sun. 
We have performed over 72$\times 10^{2}$ Milky Way pulsars simulations to account for 
i) the stochastic nature of the neutron stars' birth distribution in space and time, and uncertainties 
on the pulsars' birth rate, ii) uncertainties on the initial spin-down power distribution that Milky 
Way pulsars follow, iii) different assumptions on the evolution with time of the pulsars' spin-down 
power, iv) different assumptions on the cosmic-ray electron and positron fluxes that pulsars inject 
into the ISM and v) uncertainties on the  propagation of these cosmic rays through the ISM 
and the Heliosphere before they get detected. The range of model parameter values that we 
explore is wide and varying among different assumptions can significantly affect the observed 
electron and positron fluxes as shown in Figures~\ref{fig:Kappa_Tau0_Zeta} and~\ref{fig:ISM_Prop}.   
Each of our Milky Way pulsars
simulations contains from $5\times 10^{3}$ to $19\times 10^{3}$ unique pulsars with ages up 
to 10 Myr, depending on the exact birth rate assumption. Within a given simulation, each pulsar 
has a unique location, age, initial spin-down power and spectral index of injected cosmic rays. 
             
In performing our fits to the \textit{AMS-02}, \textit{DAMPE} and \textit{CALET} flux observations,
we account also for the existence of cosmic-ray primary electrons from SNRs and secondary 
electrons and positrons produced in inelastic collisions taking place in the ISM. We also account 
for those fluxes' respective uncertainties. Examples of our fits are given in Figures~\ref{fig:PositronsFlux},
\ref{fig:PositronFraction},~\ref{fig:TotalLeptonFlux},~\ref{fig:DampeLeptons} and~\ref{fig:CaletLeptons}. 

We find a strong preference for pulsars models with a spin-down braking index of 
$\kappa \ge 3.0$ (see Eq.~\ref{eq:SpinDown} and Figures~\ref{fig:Kappa_VS_ISM},~\ref{fig:Injec_VS_Kappa} 
and~\ref{fig:ComparisonHeatmaps}). Such a result is in contrast to observations of the about ten 
young pulsars, for which a reliable measurement of $\kappa < 3.0$ has been made (with the exception 
of one \cite{Archibald:2016hxz}). As our analysis tests the averaged properties of much older 
pulsars than the electromagnetic measurements do, our results  show that pulsars' breaking index 
evolves with time to larger values. This results in older pulsars losing their rotational energy in a slower 
manner than that predicted from the regular magnetic dipole radiation. Our results show a new way
of studying the evolution of pulsars. With higher statistics in the future we expect that specific models 
on the pulsars' braking index evolution with time can be tested. 

Furthermore, we find that pulsars inject into the ISM electrons and positrons with relatively similar 
cosmic-ray spectra that scale roughly as $dN/dE \propto E^{-1.6}$ up to $O(10)$ TeV. Also,
pulsars convert $O(10\%)$ of their rotational energy into such cosmic rays (see Figure~\ref{fig:Eta_dist}). 
Our conclusions are fairly robust to the exact birth rate of Milky Way pulsars and the exact local 
ISM assumptions. We still find a preference for larger pulsar birth rates and thicker diffusion zone
ISM models. 

Finally, when studying the lower energies we noticed that the \textit{AMS-02} positron measurements
give a spectral feature at $\simeq$12 GeV. While some of our simulations can explain such a feature,
its presence likely suggests a population of positron sources outside the volume of study, or of an entirely
different origin. We leave the possible origin of such a feature to future studies.

We have made publicly available our Milky Way pulsars simulations in their pre-fitted format for the 
entire set. We have also provided the fitted fluxes from simulations that are in agreement with the 
\textit{AMS-02}, \textit{CALET} and \textit{DAMPE} observations. These files can be found at  
\texttt{https://zenodo.org/record/5659004\#.YYqnbi-ZN0s}.
  
\textit{Acknowledgements:} 
We thank Tanvi Karwal for useful discussions and Alexandros Kehagias for establishing this collaboration.
IC acknowledges support from the NASA Michigan Space Grant Consortium, Grant No. 80NSSC20M0124.

\bibliography{Pulsar_Properties_21}

\end{document}